\documentclass{elsart}
\usepackage{graphicx}
\usepackage{epsfig}
\usepackage{amssymb}
\usepackage{amsmath}
\usepackage{latexsym}
\usepackage{cite}
\usepackage{color}
\usepackage{rotating,fancyvrb}
\begin{document}
\def\text{\textstyle}
\newcommand{\litwo}{\mbox{Li}_2}
\newcommand{\lithree}{\mbox{Li}_3}
\newcommand{\SUBR}[1]{\bigskip\fbox{\tt #1}\bigskip}
\newcommand{\ALIBABA}{{\tt ALIBABA}}
\newcommand{\ZSHAPE}{{\tt ZSHAPE}}
\newcommand{\KORALZ}{{\tt KORALZ}}
\newcommand{\ZBIZON}{{\tt ZBIZON}}
\newcommand{\BHANG}{{\tt BHANG}}
\newcommand{\DIZET}{{\tt DIZET}}
\newcommand{\dz}{{\tt DIZET}}
\newcommand{\citr}{$\circ$}
\newcommand{\nn}{\noindent}
\newcommand{\hf}{\hfill}
\newcommand{\bq}{\begin{equation}}
\newcommand{\eq}{\end{equation}}
\newcommand{\ba}{\begin{eqnarray}}
\newcommand{\ea}{\end{eqnarray}}
\newcommand{\nl}{\nonumber\\}
\newcommand{\nll}{\nonumber\\}
\newcommand{\nnn}{\\ \nonumber \\}
\newcommand{\naive}{na\"\i{}ve}
\hyphenation{brems-strah-lung}
\newcommand{\LEPI}{LEP1}
\newcommand{\LEPII}{LEP2}
\newcommand {\zf}{{\tt ZFITTER}}
\newcommand{\ee}{$e^+e^-$}
\newcommand{\oalf}{\mbox{${\cal O}(\alpha) \:$}}
\newcommand{\oalz}{\mbox{${\cal O}(\alpha^2) \:$}}
\newcommand{\os}{\mbox{${\cal O}(\alpha \alpha_s) \:$}}
\newcommand{\oaa}{\mbox{${\cal O}(\alpha^2 m_t^4) \:$}}
\newcommand{\BS}{\bigskip}
\newcommand{\Z}{$Z$}
\newcommand{\IE}{ i.e. }
\newcommand{\EG}{{ e.g. }}
\newcommand{\SWE}{$s_W^{2,{\rm{eff}}}$}
\newcommand{\Sw}{$\sin^2\theta_W$}
\newcommand{\GAMZ}{\mbox{$\Gamma_Z$}}
\newcommand{\RS}{$\sqrt{s}$}
\newcommand{\BB}{$b\bar{b}$}
\newcommand{\FF}{$f\bar{f}$}
\newcommand{\zetaz}{\mbox{$\zeta(2)$}}
\newcommand{\zetad}{\mbox{$\zeta(3)$}}
\newcommand{\rt} {\mbox{$r_{t}   $}}
\newcommand{\rtS}{\mbox{$r^2_{t} $}}
\newcommand{\rw} {\mbox{$r_{\sss{W}}  $}}
\newcommand{\rz} {\mbox{$r_{\sss{Z}}  $}}
\newcommand{\rwS}{\mbox{$r^2_{\sss{W}}$}}
\newcommand{\rzS}{\mbox{$r^2_{\sss{Z}}$}}
\newcommand{\Rw} {\mbox{$R_{\sss{W}}  $}}
\newcommand{\RwS}{\mbox{$R^2_{\sss{W}}$}}
\newcommand{\Rz} {\mbox{$R_{\sss{Z}}  $}}
\newcommand{\RzS}{\mbox{$R^2_{\sss{Z}}$}} 
\newcommand{\Rv} {\mbox{$R_{_V}  $}}
\newcommand{\RvS}{\mbox{$R^2_{_V}$}}
\newcommand{\Litwo}{\mbox{${\rm{Li}}_{2}$}}
\newcommand{\nf }{\mbox{$n_f  $}}
\newcommand{\nfS}{\mbox{$n^2_f$}}
\newcommand{\mq }{\mbox{$m_q  $}}
\newcommand{\mqS}{\mbox{$m^2_q$}}
\newcommand{\mqQ}{\mbox{$m^4_q$}}
\newcommand{\mqX}{\mbox{$m^6_q$}}
\newcommand{\mqp}{\mbox{$m'_q $}}
\newcommand{\mqpS}{\mbox{$m'^2_q$}}
\newcommand{\mqpQ}{\mbox{$m'^4_q$}}
\newcommand{\ms }{\mbox{$m_s  $}}
\newcommand{\msS}{\mbox{$m^2_s$}}
\newcommand{\mc }{\mbox{$m_c  $}}
\newcommand{\mcS}{\mbox{$m^2_c$}}
\newcommand{\mcQ}{\mbox{$m^4_c$}}
\newcommand{\mb }{\mbox{$m_b  $}}
\newcommand{\mbS}{\mbox{$m^2_b$}}
\newcommand{\mbQ}{\mbox{$m^4_b$}}
\newcommand{\Mq }{\mbox{$M_q  $}}
\newcommand{\Ms }{\mbox{$M_s  $}}
\newcommand{\Mc }{\mbox{$M_c  $}}
\newcommand{\Mb }{\mbox{$M_b  $}}
\newcommand{\Mt }{\mbox{$M_t  $}}
\newcommand{\MqS}{\mbox{$M^2_q$}}
\newcommand{\MsS}{\mbox{$M^2_s$}}
\newcommand{\McS}{\mbox{$M^2_c$}}
\newcommand{\MbS}{\mbox{$M^2_b$}}
\newcommand{\MtS}{\mbox{$M^2_t$}}
\newcommand{\MSB}{\overline{MS}}
\newcommand{\LMSB}{\mbox{$\Lambda_{\overline{\mathrm{MS}}}$}}
\newcommand{\LMSBS}{\mbox{$\Lambda^2_{\overline{\mathrm{MS}}}$}}
\newcommand{\LMSBn }{\mbox{$\Lambda^{(\nf)}_{\overline{\mathrm{MS}}}$}}
\newcommand{\LMSBnS}{\mbox{$\left(\Lambda^{(\nf)}_{\overline{\mathrm{MS}}}
\right)^2$}}
\newcommand{\LMSBnml }{\mbox{$\Lambda^{(\nf-1)}_{\overline{\mathrm{MS}}}$}}
\newcommand{\LMSBnmlS}{\mbox{$\left(\Lambda^{(\nf-1)}_{\overline{\mathrm{MS}}}
  \right)^2$}}
\newcommand{\LMSBt }{\mbox{$\Lambda^{(3)}_{\overline{\mathrm{MS}}}$}}
\newcommand{\LMSBtS}{\mbox{$\left(\Lambda^{(3)}_{\overline{\mathrm{MS}}}\right
    )^2$}}
\newcommand{\LMSBf }{\mbox{$\Lambda^{(4)}_{\overline{\mathrm{MS}}}$}}
\newcommand{\LMSBfS}{\mbox{$\left(\Lambda^{(4)}_{\overline{\mathrm{MS}}}\right)
    ^2$}}
\newcommand{\LMSBv }{\mbox{$\Lambda^{(5)}_{\overline{\mathrm{MS}}}$}}
\newcommand{\LMSBvS}{\mbox{$\left(\Lambda^{(5)}_{\overline{\mathrm{MS}}}
  \right)^2$}}
\newcommand{\als }{\alpha_{_S}}
\newcommand{\alem}{\alpha_{em}}
\newcommand{\alsS}{\alpha^2_{_S}}
\newcommand{\ztwo}{\zeta(2)}
\newcommand{\ztri}{\zeta(3)}
\newcommand{\zfor}{\zeta(4)}
\newcommand{\zfiv}{\zeta(5)}
\def\gn{\Gamma_{\nu}}
\def\ge{\Gamma_{e}}
\def\gmu{\Gamma_{\mu}}
\def\gt{\Gamma_{\tau}}
\def\gl{\Gamma_{l}}
\def\gu{\Gamma_{u}}
\def\gd{\Gamma_{d}}
\def\gc{\Gamma_{c}}
\def\gs{\Gamma_{s}}
\def\gb{\Gamma_{b}}
\def\gz{\Gamma_{\sss{Z}}}
\def\gh{\Gamma_{h}}
\def\gi{\Gamma_{\rm{inv}}}        
\def\afb{\mbox{$A_{_{FB}}$}}
\def\barf{\overline f}
\def\barq{\overline q}
\def\barb{\overline b}
\def\bart{\overline t}
\def\barc{\overline c}
\def\dr{\Delta r}
\def\Szg{\Sigma_{_{Z\gamma}}}
\def\Szz{\Sigma_{_{ZZ}}}
\def\Sww{\Sigma_{_{WW}}}
\def\Swwg{\Sigma_{_{WW}}^{^G}}
\def\Stg{\Sigma_{_{3Q}}}
\def\Stt{\Sigma_{_{33}}}
\def\Pgg{\Pi_{\gamma\gamma}}
\def\gfd{\gamma_5}
\def\mev{{\hbox{MeV}}}
\def\gev{{\hbox{GeV}}}
\def\srt{\sqrt{2}}
\def\xsf{\sigma_{_F}}
\def\xsb{\sigma_{_B}}
\def\chig{\chi_{\gamma}}
\def\chiz{\chi_{\sss{Z}}}
\def\s0h{\sigma^0_h}
\def\nl{\nonumber \\}
\def\vae{v_e a_e}
\def\vaf{v_f a_f}
\newcommand{\bqa}{\begin{eqnarray}}
\newcommand{\eqa}{\end{eqnarray}}
\newcommand{\ds }{\displaystyle}
\newcommand{\mpar}[1]{{\marginpar{\hbadness10000%
                     \sloppy\hfuzz10pt\boldmath\bf#1}}%
                      \typeout{marginpar: #1}\ignorespaces}
\newcommand{\gQ}{\ph{\scriptscriptstyle{Q}}}
\newcommand{\gL}{\ph{\scriptscriptstyle{L}}}
\newcommand{\gZQ}{\ph{\scriptscriptstyle{(Z)Q}}}
\newcommand{\gZL}{\ph{\scriptscriptstyle{(Z)L}}}
\newcommand{\ZQ}{\scriptscriptstyle{ZQ}}
\newcommand{\ZL}{\scriptscriptstyle{ZL}}
\newcommand{\QL}{\scriptscriptstyle{QL}}
\newcommand{\LQ}{\scriptscriptstyle{LQ}}
\newcommand{\zg}{{\scriptscriptstyle{Z}}\ph}
\newcommand{\sss}[1]{\scriptscriptstyle{#1}}
\newcommand{\Zi}{\sss{Z}}
\newcommand{\gf}{G_{\mu}}
\newcommand{\gfs}{G^2_{\mu}}
\newcommand{\ph}{\gamma}
\newcommand{\ab}{A}
\newcommand{\zb}{Z}
\newcommand{\wb}{W}
\newcommand{\hb}{H}
\newcommand{\fe}{e}
\newcommand{\fbe}{{\bar{e}}}
\newcommand{\ff}{f}
\newcommand{\ffp}{f'}
\newcommand{\fep}{e^{+}}
\newcommand{\fem}{e^{-}}
\newcommand{\fnue}{\nu_e}
\newcommand{\barl}{\overline{l}}
\newcommand{\fu}{u}
\newcommand{\fd}{d}
\newcommand{\fc}{c}
\newcommand{\fs}{s}
\newcommand{\ft}{t}
\newcommand{\fb}{b}
\newcommand{\ffb}{b}
\newcommand{\fl}{l}
\newcommand{\fq}{q}
\newcommand{\flm}{\mu}
\newcommand{\flmp}{\mu^{+}}
\newcommand{\flmm}{\mu^{-}}
\newcommand{\flt}{\tau}
\newcommand{\gap}{\lpar 1+\gamma_5\rpar}
\newcommand{\gadi}[1]{\gamma_{#1}}
\newcommand{\cff}[5]{C_{#1}\lpar #2;#3,#4,#5\rpar}    
\newcommand{\mws}{M^2_{\sss{W}}}
\newcommand{\mzs}{M^2_{\sss{Z}}}
\newcommand{\mzc}{M^3_{\sss{Z}}}
\newcommand{\mhs}{M^2_{\sss{H}}}
\newcommand{\mts}{m^2_{t}}
\newcommand{\mvs}{M^2_{_V}}
\newcommand{\mV }{M^2_{_V}}
\newcommand{\mf }{m^2_f}
\newcommand{\mfp}{m^2_{f'}}
\newcommand{\mfh}{m^2_{h}}
\newcommand{\mt }{m^2_t}
\newcommand{\mes}{m^2_e}
\newcommand{\mfpq}{m^4_{f'}}
\newcommand{\mwq }{M^4_{\sss{W}}}
\newcommand{\mzq }{M^4_{\sss{Z}}}
\newcommand{\mhq }{M^4_{\sss{H}}}
\newcommand{\mtq }{m^4_{t}}
\newcommand{\mhl }{M_{\sss{H}}}
\newcommand{\mVl }{M_{_V}}
\newcommand{\mwl }{M_{\sss{W}}}
\newcommand{\mzl }{M_{\sss{Z}}}
\newcommand{\mfl }{m_f}
\newcommand{\mtl }{m_t}
\newcommand{\mel }{m_e}
\newcommand{\mfpl}{m_{f'}}
\newcommand{\mfhl}{m_{h}}
\newcommand{\mml }{m_{\mu}}
\newcommand{\mfs }{m^2_f}
\newcommand{\mls }{m^2_l}
\newcommand{\uml}{ m   _{t}}
\newcommand{\um }{ m^2_{t} }
\newcommand{\umf}{ m^4_{t} }
\newcommand{\wml}{ M  _{\sss{W}}}
\newcommand{\zml}{ M  _{\sss{Z}}}
\newcommand{\dml}{ m   {_f}}
\newcommand{\dms}{ m^2_{_f}}
\newcommand{\hml}{ M  _{\sss{H}}}
\newcommand{\rtc}{ r^3_  t }
\newcommand{\sla}[1]{/\!\!\!#1}
\newcommand{\spd}{\partial}
\newcommand{\rws}{r^2_{\sss{W}}}
\newcommand{\rzs}{r^2_{\sss{Z}}}
\newcommand{\asums}[1]{\sum_{#1}}
\newcommand{\cf}{c_f}
\newcommand{\Nf}{N_f}     
\newcommand{\fbf}{{\bar{f}}}
\newcommand{\Vvert}{V}
\newcommand{\tpfi}{\lpar 2\pi\rpar^4\ib}
\newcommand{\zcont}[2]{z_{#1}^{#2}}
\newcommand{\Vverti}[3]{V_{#1}^{#2}\lpar{#3}\rpar}
\newcommand{\lpar}{\left(}                            
\newcommand{\rpar}{\right)}
\newcommand{\lrbr}{\left[}
\newcommand{\rrbr}{\right]}
\newcommand{\lcbr}{\left\{}
\newcommand{\rcbr}{\right\}} 
\newcommand{\ib  }{i}
\newcommand{\qf  }{Q_f  }
\newcommand{\qfs }{Q^2_f}
\newcommand{\qfc }{Q^3_f}
\newcommand{\qe  }{Q_e  }
\newcommand{\qes }{Q^2_e}
\newcommand{\qep }{Q_{e'}}
\newcommand{\qfp }{Q_{f'}}
\newcommand{\qfps}{Q^2_{f'}}
\newcommand{\qfpc}{Q^3_{f'}}
\newcommand{\gbs }{g^2}
\newcommand{\gbc }{g^3}
\newcommand{\Gverti}[3]{G_{#1}^{{#2}}\lpar{#3}\rpar}
\newcommand{\Zverti}[3]{Z_{#1}^{{#2}}\lpar{#3}\rpar}
\newcommand{\vverti}[3]{F^{#1}_{_{#2}}\lpar{#3}\rpar}
\newcommand{\vvertil}[3]{F^{#1}_{#2}\lpar{#3}\rpar}
\newcommand{\cvetri}[3]{{\cal{F}}^{#1}_{_{#2}}\lpar{#3}\rpar}
\newcommand{\cvetril}[3]{{\cal{F}}^{#1}_{#2}\lpar{#3}\rpar}
\newcommand{\hvetri}[3]{{\hat{\cal{F}}}^{#1}_{_{#2}}\lpar{#3}\rpar}
\newcommand{\averti}[3]{{\bar{F}}^{#1}_{_{#2}}\lpar{#3}\rpar}
\newcommand{\avetri}[3]{{\overline{\cal{F}}}^{#1}_{_{#2}}\lpar{#3}\rpar}
\newcommand{\hmix}[2]{{\hat{\Pi}^{{#1},F}}_{_{\zb\gamma}}\lpar{#2}\rpar}
\newcommand{\fverti}[2]{F^{#1}_{#2}}
\newcommand{\bDz}[2]{{\cal{D}}^{{#1},F}_{_{\zb}}\lpar{#2}\rpar}
\newcommand{\gadu}[1]{\gamma_{#1}}
\newcommand{\siw }{s_{\sss{W}}}           
\newcommand{\cow }{c_{\sss{W}}}
\newcommand{\siws}{s^2_{\sss{W}}}
\newcommand{\cows}{c^2_{\sss{W}}}
\newcommand{\siwc}{s^3_{\sss{W}}}
\newcommand{\cowc}{c^3_{\sss{W}}}
\newcommand{\siwf}{s^4_{\sss{W}}}
\newcommand{\cowf}{c^4_{\sss{W}}}
\newcommand{\vpa}[2]{\sigma_{#1}^{#2}}
\newcommand{\vma}[2]{\delta_{#1}^{#2}}
\newcommand{\vc }[1]{v_{#1}}
\newcommand{\ac }[1]{a_{#1}}
\newcommand{\vcs}[1]{v^2_{#1}}
\newcommand{\acs}[1]{a^2_{#1}}
\newcommand{\Rvaz}[1]{g^{#1}_{\sss{\zb}}}
\newcommand{\rvab}[1]{{\bar{g}}_{#1}}
\newcommand{\rvabs}[1]{{\bar{g}}^2_{#1}}
\newcommand{\rab }[1]{{\bar{a}}_{#1}}
\newcommand{\rvb }[1]{{\bar{v}}_{#1}}
\newcommand{\rabs}[1]{{\bar{a}}^2_{#1}}
\newcommand{\rva }[1]{g_{#1}}
\newcommand{\rvas}[1]{g^2_{#1}}
\newcommand{\Rva }[1]{G_{#1}}
\newcommand{\Rvac}[1]{G^{*}_{#1}}
\newcommand{\Rvah }[1]{{\hat{G}}_{#1}}
\newcommand{\Rvahc}[1]{{\hat{G}}^{*}_{#1}}
\newcommand{\Rvas}[1]{G^2_{#1}}
\newcommand{\vaeII}{{\bar{g}}^2_{\fe}+{\bar{a}}^2_{\fe}}
\newcommand{\vafII}{{\bar{g}}^2_{\ff}+{\bar{a}}^2_{\ff}}
\newcommand{\four}{{\bar{g}}_{\fe}{\bar{a}}_{\fe}{\bar{g}}_{\ff}{\bar{a}}_{\ff}}
\newcommand{\tcie}{I^{(3)}_e}
\newcommand{\tcif}{I^{(3)}_f}
\newcommand{\rhoe}{\rho_{\fe}}
\newcommand{\rhoi}[1]{\rho_{#1}}
\newcommand{\rhois}[1]{\rho^2_{#1}}
\newcommand{\rhopi}[1]{\rho'_{#1}}
\newcommand{\rhobi} [1]{{\bar{\rho}}_{#1}}
\newcommand{\rhohi} [1]{{\hat{\rho}}_{#1}}
\newcommand{\rhobpi}[1]{{\bar{\rho}}'_{#1}}
\newcommand{\saff}[1]{A_{#1}}             
\newcommand{\aff}[2]{A_{#1}\lpar #2\rpar}                   
\newcommand{\sbff}[1]{B_{#1}}             
\newcommand{\sfbff}[1]{B^{F}_{#1}}
\newcommand{\bff}[4]{B_{#1}\lpar #2;#3,#4\rpar}             
\newcommand{\fbff}[4]{B^{F}_{#1}\lpar #2;#3,#4\rpar}        
\newcommand{\scff}[1]{C_{#1}}             
\newcommand{\sdff}[1]{D_{#1}}                 
\newcommand{\dff}[7]{D_{#1}\lpar #2,#3;#4,#5,#6,#7\rpar}       
\newcommand{\delrho}[1]{{\Delta \rho}^{#1}}
\newcommand{\fbu}{{\overline{u}}}
\newcommand{\fbd}{{\overline{d}}}
\newcommand{\wbm}{W^{-}}
\newcommand{\wbp}{W^{+}}
\newcommand{\rts}{r^2_{\ft}}
\newcommand{\Lnrt}{\ln{\rt}}
\newcommand{\Rws}{R^2_{_{\wb}}}
\newcommand{\Rwc}{R^3_{_{\wb}}}
\newcommand{\Rzs}{R^2_{\sss{Z}} }
\newcommand{\rhw}{r_{_{\wb}}}
\newcommand{\rhz}{r_{_{\zb}}}
\newcommand{\rhzs}{r^2_{_{\zb}}}
\newcommand{\boxc}[2]{{\cal{B}}_{#1}^{#2}}
\newcommand{\hboxc}[3]{\hat{{\cal{B}}}_{#1}^{#2}\lpar{#3}\rpar}
\newcommand{\Pumoms}{U^2}
\newcommand{\Ptmoms}{T^2}
\newcommand{\Trmoms}{-s}
\newcommand{\sewti}[2]{w_{#1}^{#2}}
\newcommand{\Dz}[2]{{\cal{D}}_{_{\zb}}^{#1}\lpar{#2}\rpar}
\newcommand{\mix}[2]{{\cal{M}}_{#1}\lpar{#2}\rpar}
\newcommand{\Pzg}{\Pi_{\zg}}
\newcommand{\hDz}[2]{{\hat{\cal{D}}}^{{#1},F}_{_{\zb}}\lpar{#2}\rpar}
    \newcommand{\tHs}{\mu}
    \newcommand{\tHss}{\mu^2}
    \newcommand{\stwl}{s_{\sss{W}}  }
    \newcommand{\ctwl}{c_{\sss{W}}  }
    \newcommand{\stws}{s^2_{\sss{W}}}
    \newcommand{\stwf}{s^4_{\sss{W}}}
    \newcommand{\ctws}{c^2_{\sss{W}}}
    \newcommand{\ctwf}{c^4_{\sss{W}}}
 \newcommand{\qd }{Q_f  }
 \newcommand{\qds}{Q^2_f}
 \newcommand{\qdc}{Q^3_f}
  \newcommand{\vmad }{\delta_f  }
  \newcommand{\vmads}{\delta^2_f}
  \newcommand{\vpau }{\sigma_{f'}}
  \newcommand{\vmaes}{\delta^2_e }
  \newcommand{\zaOfer}{\Pi^{\fer,F}_{\gamma\gamma}(0)}
\newcommand{\bos}{\rm{bos}}
\newcommand{\fer}{\rm{fer}}
\newcommand{\pole}{\left( \dlt-\Lnw \right)}
\newcommand{\deltar}{ \Delta r}
\newcommand{\Trqf}{ \sum_f c_f Q^2_f }
\newcommand{\qum}{ \left| Q_{f'} \right| }
\newcommand{\qdm}{ \left| Q_{f}  \right| }
\newcommand{\zmz}{\Sigma^{'}_{_{\zb\zb}}(\mzs)}                          
\newcommand{\wwUf}{w^F_{_{W}}}
\newcommand{\Lnw }{\ln \frac{\mws}{\mu^2}}   
\newcommand{\Lnfw}{\ln \frac{\dms}{\mws} }
\newcommand{\Lnew}{\ln \frac{\mes}{\mws} }
\newcommand{\Lns }{\ln \frac{\um}{\mu^2}}
\newcommand{\au }{a_{f'}}
\newcommand{\vd }{v_f}
\newcommand{\ad }{a_f}
\newcommand{\ve }{v_e}  
\newcommand{\tvpad}{(3v^2_f +a^2_f)}
\newcommand{\tvpae}{(3 v^2_e +a^2_e)}
\newcommand{\vpad }{\sigma_f}
\newcommand{\vpae }{\sigma_e}
\newcommand{\vpadpa}{\sigma^a_{f^{'}}}
\newcommand{\vpada }{\sigma^a_f}
\newcommand{\vpadae}{\sigma^a_e}
\newcommand{\pd}[1]{\partial_{#1}}
\newcommand{\asum}[3]{\sum_{#1=#2}^{#3}}
\newcommand{\lkall}[3]{\lambda\lpar#1,#2,#3\rpar}       
\newcommand{\mind}[1]{m_{#1}}
\newcommand{\minds}[1]{m^2_{#1}}
\newcommand{\mindc}[1]{m^3_{#1}}
\newcommand{\mindf}[1]{m^4_{#1}}
\newcommand{\Mind}[1]{M_{#1}}
\newcommand{\Minds}[1]{M^2_{#1}}
\newcommand{\Mindc}[1]{M^3_{#1}}
\newcommand{\Mindf}[1]{M^4_{#1}}
\newcommand{\imom}{q}
\newcommand{\imomi}[1]{q_{#1}}
\newcommand{\imoms}{q^2}
\newcommand{\pmom}{p}
\newcommand{\pmomp}{p'}
\newcommand{\pmoms}{p^2}
\newcommand{\pmomq}{p^4}
\newcommand{\pmomx}{p^6}
\newcommand{\pmomi}[1]{p_{#1}}
\newcommand{\pmomis}[1]{p^2_{#1}}
\newcommand{\dlt}{\displaystyle{\frac{1}{\epsb}}}
\newcommand{\epsh}{\hat\varepsilon}
\newcommand{\epsb}{\bar\varepsilon}
\newcommand{\Ddrh}{{\ds\frac{1}{\hat{\varepsilon}}}}
\newcommand{\ep}{\epsilon}
\newcommand{\chapt}[1]{Chapter~\ref{#1}}
\newcommand{\chaptsc}[2]{Chapter~\ref{#1} and \ref{#2}}
\newcommand{\eqn}[1]{Eq.~(\ref{#1})}
\newcommand{\eqns}[2]{Eqs.~(\ref{#1})--(\ref{#2})}
\newcommand{\eqnss}[1]{Eqs.~(\ref{#1})}
\newcommand{\eqnsc}[2]{Eqs.~(\ref{#1}) and (\ref{#2})}
\newcommand{\eqnst}[3]{Eqs.~(\ref{#1}), (\ref{#2}) and (\ref{#3})}
\newcommand{\eqnsf}[4]{Eqs.~(\ref{#1}), 
          (\ref{#2}), (\ref{#3}) and (\ref{#4})}
\newcommand{\eqnsv}[5]{Eqs.(\ref{#1}), 
          (\ref{#2}), (\ref{#3}), (\ref{#4}) and (\ref{#5})}
\newcommand{\tbn}[1]{Tab.~\ref{#1}}
\newcommand{\tabn}[1]{Tab.~\ref{#1}}
\newcommand{\tbns}[2]{Tabs.~\ref{#1}--\ref{#2}}
\newcommand{\tabns}[2]{Tabs.~\ref{#1}--\ref{#2}}
\newcommand{\tbnsc}[2]{Tabs.~\ref{#1} and \ref{#2}}
\newcommand{\fig}[1]{Fig.~\ref{#1}}
\newcommand{\figs}[2]{Figs.~\ref{#1}--\ref{#2}}
\newcommand{\figsc}[2]{Figs.~\ref{#1} and \ref{#2}}
\newcommand{\sect}[1]{Section~\ref{#1}}
\newcommand{\sects}[2]{Sections~\ref{#1} and \ref{#2}}
\newcommand{\subsect}[1]{Subsection~\ref{#1}}
\newcommand{\appendx}[1]{Appendix~\ref{#1}}
\newcommand{\sman}{s}
\newcommand{\tman}{t}
\newcommand{\uman}{u}
\newcommand{\smani}[1]{s_{#1}}
\newcommand{\smanp}{s'}
\newcommand{\bsmani}[1]{{\bar{s}}_{#1}}
\newcommand{\smans}{s^2}
\newcommand{\tmans}{t^2}
\newcommand{\umans}{u^2}
\newcommand{\gspi}{\frac{g^2}{16\pi^2}}
\newcommand{\drrem}{\deltar_{\rm rem}}
\newcommand{\deltarremho}{\deltar^{ho}_{\rm rem}}
\newcommand{\dalpha}{\Delta\alpha} 
\newcommand{\dalphav}{\Delta\alpha^{(5)}(\mzs)} 
\newcommand{\dalphat}{\Delta\alpha^{\ft}(\mzs)} 
\newcommand{\Reb}{{\rm{Re}}}
\newcommand{\Imb}{{\rm{Im}}}
\newcommand{\pir}[1]{\Pi^{\rm{\sss{R}}}\lpar #1\rpar}
\newcommand{\mqs}{m^2_{q}}
\newcommand{\ord}[1]{{\cal O}\lpar#1\rpar}
\newcommand{\prot}{p}
\newcommand{\aprot}{{\bar{p}}}
\newcommand{\Nucln}{N}
\newcommand{\dalhv}{\Delta\alpha^{(5)}_{\had}(\mzs)}
\newcommand{\dall}{\Delta\alpha_{\lep}}
\newcommand{\lep}{{l}}
\newcommand{\had}{{h}}
\newcommand{\drho}{\Delta\rho}
\newcommand{\drhov}{\delta\rho}
\newcommand{\dkapv}{\delta\kappa}
\newcommand{\drhovh}{\delta{\hat{\rho}}}
\newcommand{\drhovb}{\delta{\hat{\rho}}}
\newcommand{\EW}{\sss{\rm{EW}}}
\newcommand{\seffsf}[1]{\sin^2\theta^{#1}_{\rm{eff}}}
\newcommand{\drh}{\Delta{\hat{r}}}
\newcommand{\rZf}{\rho^{\ff}_{\sss{Z}}}
\newcommand{\rZl}{\rho^{\fl}_{\sss{Z}}}
\newcommand{\gdp}{\gamma_{+}}
\newcommand{\kZf}{\kappa^{\ff}_{\sss{Z}}}
\newcommand{\rZdf}[1]{\rho^{#1}_{\sss{Z}}}
\newcommand{\kZdf}[1]{\kappa^{#1}_{\sss{Z}}}
\newcommand{\wt}{ w_t}
\newcommand{\zt}{ z_t}
\newcommand{\Ht}{ h_t}
\newcommand{\xts}{x^2_t}
%
%
\newcommand{\gel}{\Gamma_{\fe}}
\newcommand{\gff}{\Gamma_{\ff}}
\newcommand{\gll}{\Gamma_{\fl}}
\newcommand{\gqq}{\Gamma_{\fq}}
\newcommand{\gq}{\Gamma_{\fq}}
\newcommand{\gbq}{\Gamma_{\fb}}
\newcommand{\gzs}{\Gamma^2_{\sss{Z}}}
\newcommand{\drhigs }{\deltar^{H}_{\rm{res}}}
\newcommand{\drhigsa}{\deltar^{H,\alpha}_{\rm{res}}}
\newcommand{\drhigsg}{\deltar^{H,G}_{\rm{res}}}
\newcommand{\Ksc}{K_{\rm{scale}}}
\newcommand{\afba}[1]{A^{#1}_{_{\rm FB}}}
\newcommand{\alra}[1]{A^{#1}_{_{\rm LR}}}
\newcommand{\Pzga}[2]{\Pi^{#1}_{_{\zb\ab}}\lpar#2\rpar}
\newcommand{\vfwi}[1]{\sigma^{a}_{#1}}
\newcommand{\vfwsi}[1]{\lpar\sigma^{a}_{#1}\rpar^2}
\newcommand{\qb}{Q_b}
\newcommand{\qt}{Q_t}
\newcommand{\qus}{Q^2_u}
\newcommand{\qbs}{Q^2_b}
\newcommand{\qts}{Q^2_t}
\newcommand{\xvar}{x}
\newcommand{\xvars}{x^2}
\newcommand{\rvar}{r}
\newcommand{\rvari}[1]{r_{#1}}
\newcommand{\vf}{(v^2_f+a^2_f)}
\newcommand{\vb}{\bar{v}}
\newcommand{\lpoli}[1]{\lambda_{#1}}
\newcommand{\hpoli}[1]{h_{#1}}
\newcommand{\reni}[1]{R_{#1}}
\newcommand{\renis}[1]{R^2_{#1}}
\newcommand{\sreni}[1]{\sqrt{R_{#1}}}
\newcommand{\kappai}[1]{\kappa_{#1}}
\newcommand{\Imsi}[1]{I^2_{#1}}
\newcommand{\bgz}[1]{{\overline{\Gamma}}_{#1}}
\newcommand{\pgz}[1]{\Gamma_{#1}}
\newcommand{\intfx}[1]{\int_{\scriptstyle 0}^{\scriptstyle 1}\,d#1}
\newcommand{\intmomi}[2]{\int\,d^{#1}#2}

\def\theequation{\arabic{section}.\arabic{equation}}
\def\thetable{\arabic{section}.\arabic{table}}
\newcommand{\eqnzero}{\setcounter{equation}{0}}
\newcommand{\cal}{\mathcal}
\begin{frontmatter}
\begin{flushleft}
\footnotesize{ 
DESY 05--034
\\
UCD-EXPH/050701
\\
FERMILAB-Pub-05-256-T 
\\
WUE-ITP-2005-004
\\
SFB/CPP--05--22
}
\end{flushleft}

\title{\zf: a semi-analytical program for 
fermion pair production in  \ee\  annihilation,
from version 6.21 to version 6.42}

\author[ad7]{A. B. Arbuzov},
\author[ad1,ad2]{M. Awramik},
\author[ad3,ad4]{M. Czakon},
\author[ad5]{A. Freitas},
\author[ad6]{M. W. Gr\"unewald},
\author[ad1]{K. M\"onig},
\author[ad1]{S. Riemann},
\author[ad1]{T. Riemann\corauthref{TR}}

\corauth[TR]{Corresponding author. E-mail address: tord.riemann@desy.de}

\address[ad7]{Bogoliubov Laboratory of Theoretical Physics, JINR,
  Dubna, 141980, Russia}
\address[ad1]{DESY, Platanenallee 6, D-15738 Zeuthen, Germany}
\address[ad2]{Institute of Nuclear Physics PAS, Radzikowskiego 152, PL-31342 Cracow, Poland}
\address[ad3]{Institut f\"ur Theoretische Physik und Astrophysik, Universit\"at W\"urzburg, Am Hubland, D-97074 W\"urzburg, Germany}
\address[ad4]{Institute of Physics, University of Silesia, Uniwersytecka 4, PL-40007 Katowice, Poland}
\address[ad5]{Theoretical Physics Division, Fermilab, P. O. Box 500, Batavia, IL 60510, USA}
\address[ad6]{Department of Experimental Physics, University College Dublin, Dublin 4, Ireland}

\begin{abstract}
\zf\ is a Fortran program for the calculation of fermion pair
production and radiative corrections at high energy $e^+e^-$ colliders;
it is also suitable for
other applications where electroweak radiative corrections appear.
\zf\ is based on a semi-analytical approach to the calculation of 
radiative corrections in the Standard Model.
We present a summary of new features of the \zf\ program version 6.42
compared to version 6.21. 
The most important additions are:
(i) some higher-order QED corrections to fermion pair production,
(ii) electroweak one-loop corrections to atomic parity violation,
(iii) electroweak one-loop corrections to ${\bar \nu}_e \nu_e$ production,
(iv) electroweak two-loop corrections to the $W$ boson mass
and the effective weak mixing angle. 
\end{abstract}

\end{frontmatter}

\clearpage
\tableofcontents
\newpage
\listoftables
\newpage
\eqnzero
\addcontentsline{toc}{section}{New version summary}
\section*{New version summary}
\label{ZFITTER}
\underline{Title of program:}  \hspace{0.5cm}  
{\zf\ version 6.42 (18 May 2005)
} \\[.3cm]
\underline{Authors of original program:}       
\hspace{0.5cm}
{D.~Bardin, 
P.~Christova, 
M.~Jack,
L.~Kalinovskaya,
A.~Olchevski, 
S.~Riemann, 
T.~Riemann}
\\[.3cm]
\underline{Program obtainable from:} 
\\
{http://www-zeuthen.desy.de/theory/research/zfitter/} (main web site),
\\
{/afs/cern.ch/user/b/bardindy/public/ZF6\_42}
\\[.3cm]
\underline{Reference for \zf\ version 6.21} 
\hspace{0.5cm}
D.~Bardin et al., {Comput. Phys. Commun.} {133} (2001) 229--395
\\[.3cm]
\underline{Operating system:} 
\hspace{0.5cm}
{\tt UNIX/LINUX}, program tested under, e.g., {\tt HP-UX} and {\tt PC/Linux}
\\[.3cm]
\underline{Programming language used:} 
\hspace{0.5cm}
{\tt FORTRAN 77}
\\[.3cm]
\underline{High speed storage required:} 
\hspace{0.5cm}
 $<$ 2 MB
\\[.3cm]
\underline{No. of cards in combined program and test deck:} 
\hspace{0.5cm}
about 26,200
\\[.3cm]
\underline{Keywords:} 
\hspace{0.5cm}
Quantum electrodynamics (QED), Standard Model, electroweak interactions,
heavy boson $Z$, $e^+e^-$-annihilation, fermion pair production,
radiative corrections, initial state radiation (ISR),
final state radiation (FSR), QED interference,
SLD, LEP, ILC.
\\[.3cm]
\underline{Nature of the physical problem:} 
\hspace{0.5cm}
Fermion pair production is an important reaction for precision tests
of the Standard Model, at LEP/SLC and future linear colliders at higher 
energies.
For this purpose, QED, electroweak and QCD radiative corrections have
to be calculated 
with high precision, including higher order
effects.
Multi parameter fits used to extract model parameters from experimental
measurements require a program of sufficient flexibility and high 
calculational speed.
\zf\ combines these two aspects by employing analytical integrations
of matrix elements and at most one one dimensional numerical integration, 
as well as a variety of flags defining the physics content used.
The calculated predictions are typically at the per mille
precision level, sometimes better.
\\[.3cm]
\underline{Method of solution:} 
\hspace{0.5cm}
Numerical integration of analytical formulae.
\\[.3cm]
\underline{Restrictions on the complexity of the problem:} 
\hspace{0.5cm}
Fermion pair production is described below the top quark pair production
threshold.
Photonic corrections are taken into account with simple
cuts on photon energy, or the energies and
acollinearity of the two fermions, and {\em one} fermion production angle.
The treatment of Bhabha scattering is less advanced.
\\[.3cm]
\underline{Typical running time:} 
\hspace{0.5cm}
On a Pentium IV PC installation (2.8 GHz) using g77 under Linux 2.4.21,
approximately 23 sec are needed to run the standard test of subroutine
{\tt ZFTEST}.
This result is for a {\em default/recommended}
setting of the input parameters, with {\em all} corrections in the
Standard Model switched {\em on}.
{\tt ZFTEST} computes 12 cross-sections and cross-section asymmetries
for 8 energies with 5 interfaces, i.e. about 360 cross-sections in 23
seconds.

\clearpage
\eqnzero

\newpage

\section{Introduction\label{intro}}
The Fortran program \zf\ is based on a semi-analytical
approach to the calculation of fermion pair production in \ee\ 
annihilation at a wide range of centre-of-mass energies, 
including SLC/LEP1, LEP2, and ILC energies below the $t\overline{t}$-threshold.
\zf\ allows the calculation of several quantities needed for precision
studies of the Standard Model:
\begin{itemize}
\item $M_W$ -- the $W$ boson mass;
\item
$\Gamma_{Z}$, $\Gamma_{W}$,  --  total (and also partial) $Z$ and $W$ boson
decay widths; 
\item
${d\sigma}/{d\cos\vartheta}$ -- differential cross-sections; 
\item
$\sigma_T$ -- total cross-sections;
\item
$A_{FB}$ -- forward-backward asymmetries;
\item
$A_{LR}$ -- left-right asymmetries;
\item
$A_{pol}, A_{FB}^{pol}$ -- final state polarisation effects for $\tau$
leptons.
\end{itemize}
All observables are calculated including radiative corrections using
the (running) fine
structure constant $\alpha$, the muon decay constant $G_{\mu}$, the $Z$
mass
$M_Z$ as well as the fermion masses and the Higgs mass $M_H$ as input.

\zf\ version 6.21 and higher was mainly intended for the use at LEP1/SLC 
energies, and
versions since 6.30 are also adapted to the LEP2
kinematics. 
Various interfacing subroutines (short: interfaces) allow the user to 
calculate observables, e.g., for fits to the
experimental data with different sets of free parameters.

The Fortran package {\tt DIZET}, a library for the calculation of
electroweak radiative corrections, is part of the \zf\ distribution.
It can also be used in a stand-alone mode.
On default, {\tt DIZET} performs the following calculations:
\begin{itemize}
\item
by call of subroutine {\tt DIZET}: $W$ mass, $Z$ and $W$ partial and total
decay widths;
\item
by call of subroutine {\tt ROKANC}: four weak neutral-current (NC) 
form factors, running electromagnetic 
 and strong couplings needed for the calculation of effective NC Born 
cross sections for the production of massless fermions (however, the mass
of the top quark appearing in the virtual state of loop diagrams
for the process $e^+ e^- \to f \bar{f}$ is not ignored); 
\item
by call of subroutine {\tt RHOCC}: the corresponding form factors and running 
strong coupling for the calculation of effective CC Born cross sections; 
\item
by call of subroutine {\tt ZU\_APV}: $Q_W(Z,A)$ -- the weak charge used for
the description of parity violation in heavy atoms.
\end{itemize}
If needed, the form factors of cross sections may be made to contain
the contributions from  
$WW$ and $ZZ$ box diagrams thus ensuring  the correct kinematic
behaviour over a larger energy range compared to the Z pole.

\bigskip

\zf\ version 6.21 \cite{zfitter:v6.21new} was released in July 1999 and
was described in \cite{Bardin:1999yd}.   
Since then, there were several important developments of the program.
The current release is that of \zf\ version 6.42, dated 18 May 2005.
This article describes the changes and additions in \zf\ from version 6.21 
to version 6.42.  
Other sources of information on \zf\ and its use are
given by the previous program description and the references therein, 
the \zf\ webpage \cite{zfitter-support-page}, 
the studies \cite{Bardin:1999gt,Bardin:1999ak}, the 1999/2000 CERN
LEP2 workshop proceedings  
\cite{Kobel:2000aw}, and the studies of the LEPEWGG (LEP
electroweak working group) \cite{LEPEWWG:2005aa}.  

The essential changes from \zf\ v.~6.21 to v.~6.42 in terms of 
physics topics are: 
\begin{itemize}
\item
Higher order QED corrections to fermion-pair production, of importance
at energies off the $Z$ boson peak;
\item
Electroweak corrections to the weak charge $Q_W$, describing the
parity violation effects in atoms,  of importance for so-called global
Standard Model fits;
\item
Electroweak corrections to ${\bar \nu}_e \nu_e$ production,  of importance
for a precise description of ${\bar \nu} \nu \gamma$ production;
\item
Electroweak two-loop corrections to $M_W$ and the effective weak
mixing angle $\sin^{2,eff}\theta_W$,  of importance for global
Standard Model fits and for precise predictions of the Higgs mass $M_H$. 
\end{itemize}
Further, an option to change the strength of the Wtb vertex, $|V_{tb}|$,
was implemented in {\tt DIZET} since version 6.30: $|V_{tb}|$ is now one of
the parameters 
of subroutine {\tt DIZET}. The default is the Standard Model calculation
with $|V_{tb}|=1$. 
Several new interface routines deal with this case,
see Appendix \ref{interfaces} for the details.

This program description update is organized as follows.
In Sections \ref{sec-qed} to \ref{changes}, we describe the
improved or new physics issues.
Section \ref{input} contains a description of the input parameters and
pseudo observables, and Section \ref{zftest} reproduces reference outputs from
a running of the sample program package.
A Summary closes this update note. 
In Appendices \ref{dizetug} to \ref{interfaces} we collect, for the convenience
of the user, updated \zf\ and \dz\ user guides and an overview 
of the user interfaces.
The presentation assumes, of course, a familiarity of the reader with
\cite{Bardin:1999yd}. 

With the present update of \zf\ and its description, the
maintenance of the program has been migrated to a 
group of volunteers, the \zf\ support group.
The main webpage has been migrated to: \\
{http://www-zeuthen.desy.de/theory/research/zfitter/}
\\
The decision to do so was taken together with the original authors of the
program.
Since we did not want to change the list of authors in view of the 
long-standing history of the program, the idea to create a
\zf\ support group was considered to be an appropriate way to
handle the maintenance situation requiring permanent care 
about the program.
The \zf\ support group is composed of authors of
\zf\ version 6.21, long-term users of the program, and colleagues who
contributed substantially to its present state. 
We hope that a kernel of us will stay with \zf\ as long as the code
is needed by the community.

A correct citation of the \zf\ package will include reference
\cite{Bardin:1999yd}, together with the present update. 

 
\eqnzero
\section{Higher order QED corrections\label{sec-qed}}
Measurements at LEP1/SLC were performed in the vicinity of the $Z$
resonance so that many of the photonic corrections were suppressed.
Cross sections at energies away from the $Z$ resonance peak have a
much stronger dependence on higher order photonic corrections and
their inclusion is of numerical importance at LEP2.
Several improvements in this respect have been performed since \zf\
version 6.21 and will be described in the following subsections.
They have also been discussed in \cite{Kobel:2000aw}.

\subsection{\label{sec-qed-1}Second order initial state fermion pair
  corrections} 
Since \zf\ version 6.21 there is an improved treatment of second order
corrections for angular distributions and $A_{FB}$.
These corrections were applied 
in the leading logarithmic approximation
as described in Ref.~\cite{Bardin:1999yd}.
The option is accessed by a new flag:
\\
{\tt FBHO} = 0 -- old treatment with photonic corrections only,
\\
{\tt FBHO} = 1 -- leading log fermion pair corrections are added.
\\  
The non-singlet and singlet fermion pair contributions to the electron
(positron) structure function of \cite{Skrzypek:1992vk} (see Eq. (11)
in \cite{Arbuzov:1999cq} and Eq.~(47) in \cite{Skrzypek:1992vk})
are used 
(with different options according to the
values of the flags {\tt ISPP}, {\tt IPFC} and {\tt IPSC}).
They are directly added to the corresponding photonic 
contributions~\cite{Bardin:1989cw,Beenakker:1989km}, 
which are governed by the flag {\tt FUNA}.
For pair corrections in general, the new value of 
flag ${\tt IPTO}=-1$ is added. It allows to calculate pure 
virtual pair contributions separately. 
This option can be used for comparisons or in the case, when the
contribution of real pair emission is taken from another program.

\subsection{\label{sec-qed-2}Second order initial state QED
  corrections with acceptance 
cuts}
\newcommand{\dd}{{\mathrm d}}
Since \zf\ version 6.30,
the second order initial state radiation (ISR) QED corrections in presence
of angular cuts are improved \cite{Arbuzov:2001rt}.
A new option governed by a new flag {\tt FUNA} is implemented, with:
\\
{\tt FUNA}=0 -- old treatment,
\\
{\tt FUNA}=1 -- new treatment.
\\  
The corrected treatment of these corrections takes the angular
acceptance cuts {\tt ANG0}, {\tt ANG1} into account.
 The corrections are relevant for 
the angular distribution and for the integrated forward-backward asymmetry. 
We use the leading logarithmic
approximation by means of the electron structure function 
formalism~\cite{Skrzypek:1992vk,Arbuzov:1999cq}.
In fact the differential angular distribution of the electron--positron
annihilation process can be represented in a form analogous to that
of the Drell--Yan process: 
\begin{eqnarray} \label{a:dsigdc}
&& \frac{\dd\sigma(s,c)}{\dd c} = \int_0^1\dd x_1 {\mathcal D}(x_1,L_e)
\int_0^1 \dd x_2 {\mathcal D}(x_2,L_e)
\frac{\dd\hat\sigma(\hat{s},\hat{c})}{\dd \hat{c}} {\mathcal J} 
\Theta(\hat{s} - s'), 
\end{eqnarray}
with $c = \cos \vartheta$ and 
\begin{eqnarray}
\hat{s} &=& x_1x_2s, 
\\ 
{\mathcal J} &=& \frac{4x_1x_2}{[x_1+x_2-c(x_1-x_2)]^2},
\end{eqnarray}
where the structure functions ${\mathcal D}(x_{i},L_e)$ give the probability
to find an electron (positron) with a reduced energy fraction $x_{1(2)}$ in the initial
electron (positron). The Born--level annihilation cross section 
$\dd\hat\sigma(\hat{s},c)/\dd \hat{c}$ is defined in the 
center--of--mass reference frame of electron and positron with reduced
energy fractions. The Jacobian ${\mathcal J}$ is coming from the relation 
to the angles in the laboratory reference frame.
The structure functions are taken in the leading logarithmic
approximation keeping the first and second order photonic contributions:
\begin{eqnarray} 
{\mathcal D}(x,L_e) &=& \delta(1-x) + \frac{\alpha}{2\pi}(L_e-1)
P^{(1)}(x) +  \frac{1}{2}\left(\frac{\alpha}{2\pi}(L_e-1)\right)^2 P^{(2)}(x),
\end{eqnarray}
with
\begin{eqnarray} 
L_e &\equiv& \ln\frac{s}{m_e^2},
\\
P^{(1)}(x) &=&  \lim_{\Delta\to 0}\biggl\{ \delta(1-x)\biggl(
2\ln\Delta + \frac{3}{2} \biggr) +  \Theta(1-\Delta -x)
\frac{1+x^2}{1-x} \biggr\},  
\\
P^{(2)}(x) &=& \lim_{\Delta\to 0}\biggl\{ \delta(1-x)\biggl[
\biggl(2\ln\Delta + \frac{3}{2}\biggr)^2 - \frac{2\pi^2}{3} \biggr]
\\ \nonumber
&&+~ \Theta(1-\Delta -x)2\biggl[ \frac{1+x^2}{1-x}\biggl(
2\ln(1-x)-\ln x + \frac{3}{2}\biggr) 
\\ \nonumber
&&+~ \frac{1+x}{2}\ln x - 1 + x
\biggr] \biggr\}.
\end{eqnarray}
In order to avoid a double counting, we expanded formula~(\ref{a:dsigdc}) in
$\alpha$ and take only the ${\mathcal O}(\alpha^2)$ terms, which are
then added to the full first order corrections. Where possible, we
performed integrations over the angle and the energy fractions analytically.
This allows to get a relatively fast code for these corrections.
The calculation is realized with subroutine {\tt funang.f}.
It is compatible with the use of {\tt ICUT} = 1,2,3. 
We just mention that for the contribution to the total cross section the
complete ${\mathcal O}(\alpha^2)$ formulae are used.

\subsection{\label{sec-qed-3}Higher order photonic corrections from
  the initial-final state 
  interference}
The exponentiation of photonic initial-final-state interference
corrections was implemented according to Ref.~\cite{Greco:1975wq}.
It allows to take into account the most significant part 
of higher order corrections coming from the initial-final state 
interference. 
The corrections are relevant for 
the angular distribution and for the integrated forward-backward asymmetry. 
A combination of the one--loop initial--final interference corrections
with the  
corresponding higher order effects from the exponentiation can be
computed in \zf\ by using flag {\tt INTF} = 2. The old options are:
\\
{\tt INTF} = 0: the initial-final state interference in photonic corrections
is omitted;
\\ 
{\tt INTF} = 1: the interference is taken in the one-loop approximation. \\
The numerical effect of the exponentiation was discussed in 
Ref.~\cite{Kobel:2000aw}. It was found to be close 
to the one obtained by a slightly different exponentiation procedure
~\cite{Jadach:2000ir}.
  
\subsection{\label{sec-qed-4}Final state fermion pair production corrections}
Since \zf\ version 6.30, the corrections from
final state radiation (FSR) of fermion pairs are
implemented \cite{Arbuzov:2001rt}, according 
to the formulae given in \cite{Hoang:1995ht}. 
The option is governed by a new flag:     \\
{\tt FSPP} = 0 -- without FSR pairs,        \\
{\tt FSPP} = 1 -- with FSR pairs, additive, \\
{\tt FSPP} = 2 -- with FSR pairs, multiplicative. \\
For the best approximation, the FSR pair corrections should be treated 
multiplicative ({\tt FSPP} = 2) with respect to the ISR photonic corrections
(see Eq.~(2.1) in \cite{Arbuzov:2001rt}). 
The additive treatment of FSR pairs ({\tt FSPP} = 1) can be used for a
comparison. 
For the FSPP corrections, a cut on the invariant mass of the secondary pair
is accessible.
In order to accommodate this cut value, the variable {\tt SIPP} of the \\
{\tt SUBROUTINE ZUCUTS(INDF,ICUT,ACOL,EMIN,S$\_$PR,ANG0,ANG1,SIPP)}\\
is now used. 
The dependence on this cut for realistic event selections
was shown to be rather weak~\cite{Arbuzov:2001rt}.
Therefore, the meaning of variable {\tt SIPP} has been changed.        
Now it has nothing to do with cutting of the initial state pairs. 
In fact, there is no possibility to directly cut the initial state secondary 
pairs. 
The primary pair invariant mass cut ({\tt S$\_$PR}) is taken
into account in the phase space of secondary pairs.

\eqnzero
\section{\label{sec-nunu}Improved Born approximation for $e^+e^- \to {\bar
    \nu}\nu$ in \zf\ version 6.34}
For a study of the reaction
\ba
e^+e^- \to {\bar \nu}\nu (n\gamma),~~~~\nu=\nu_e, \nu_{\mu}, \nu_{\tau}
\ea 
one needs the effective Born approximation for
\ba
\label{eq-dsig}
\frac{d\sigma}{d\cos\vartheta} &=& \sum_{i=e,\mu,\tau}
\frac{d\sigma(e^+ e^- \to {\bar \nu}_i \nu_i)}{d\cos\vartheta} 
= 
~~ 3~ \sigma_s + \sigma_{st} + \sigma_t.
\label{eq-int}
\ea
The specific property of this reaction is due to the interference of $s$ channel $Z$
boson and $t$ channel $W$ boson exchange for $ {\bar \nu_e}\nu_e$ production.
For an application in KKMC \cite{Jadach:1999vf}, the corresponding
formulae have 
been derived \cite{Bardin:2001vt} from a related study
\cite{Bardin:1989vz,Bardin:1987rz} and implemented 
in \zf\ version 6.34 (05 Feb 2001). 
The changes in the program consist of a modification of the weak
charged current form factor (variable {\tt ROW}) in subroutine {\tt
  coscut} in {\tt zfbib6\_34.f}. 
The addition is \footnote{In the Fortran code until version 6.42, the
  factor of 4 at the r.h.s. of Equation (\ref{delta_CC-NC}) is lacking.}:
\begin{equation}
\label{delta_CC-NC}
\delta_{\tt CC-NC}={\tt QED\_CC} - {\tt QED\_NC}= \frac{\alpha}{2\pi} Q_e^2 
\left[
\frac{3}{2} \ln \frac{M_W^2}{s} + \frac{1}{2} \ln^2\frac{t}{s} 
-4~ {\rm Li}_2(1)+2
\right].
\end{equation}
The variables {\tt SIGST} and  {\tt SIGT} are influenced by this, they
correspond to $\sigma_{st}$ and $\sigma_t$.
Depending on the setting of flag {\tt ENUE} there, part or all of
them contribute to the prediction of ${d\sigma}/{d\cos\vartheta}$; see
also Section \ref{zuflag}.

In \zf\ version 6.42, the file {\tt zfbib6\_34.f} is replaced by  {\tt
  zfbib6\_40.f}.
The latter file contains a subroutine  {\tt coscut} where the
described features are not accessible.
The file {\tt zfEENN\_34.f} is no longer part of the distribution.

\eqnzero
\section{Atomic parity violation\label{sect:apv}} 
The global precision tests of the Standard Model may include an
experimental input from atomic parity violation measurements in heavy
atoms. 
The observable quantity of interest is the so-called weak charge $Q_W$: 
\ba
Q_{W}(Z,A) = 
- 2\left[  \left(2Z+N\right) C_{1u} + \left(Z+2N\right)
  C_{1d}\right].
\ea
Here, $Z$ and $N$ are the numbers of protons and neutrons in the nucleus. 
The weak couplings involved are those which parameterize the
electron-quark parity-violating Hamiltonian at zero momentum transfer:
\bqa
H_{{PV}} &=& 
\frac{G_{F}}{\sqrt 2}
(
   C_{1u}{\bar e} \gamma_{\mu}\gamma_5 e {\bar u} \gamma_{\mu} u
 + C_{2u}{\bar e} \gamma_{\mu}     e {\bar u} \gamma_{\mu} \gamma_5 u
 + C_{1d}{\bar e} \gamma_{\mu}\gamma_5 e {\bar d} \gamma_{\mu} d
\nl && 
 +~ C_{2d}{\bar e} \gamma_{\mu}     e {\bar d} \gamma_{\mu} \gamma_5 d
).
\label{APV_MS}
\eqa
In Born approximation, $ C_{1u} = -\frac{1}{2}(1-8/3 \sin^2\theta_W) $
and $ C_{1d} =  \frac{1}{2}(1-4/3 \sin^2\theta_W)$;
generally, $C_{1q} = 2 a_e v_q$ and  $C_{2q} = 2 v_e a_q$, $q=u,d$.  
In \cite{Bardin:2001ii}, using the results of \cite{Andonov:2002xc},
the higher order 
corrections to atomic parity violation in 
the on-mass-shell renormalization scheme have been derived and the
corresponding expressions were used for \zf\ version 6.34 (26 Jan 2001)
onwards.
The formulae given in \cite{Marciano:1982mm} were also reproduced. 

In subroutine {\tt ZU\_APV} the corresponding expressions 
for $C_{1u}, C_{1d}, C_{2u}, C_{2d}$ 
are calculated. In fact, the sign conventions have been changed
compared to \cite{Bardin:2001ii} in order to share the definitions of
the particle data group \footnote{We just mention that the signs of
  Equations (4) in 
  \cite{Bardin:2001ii} are not in accordance with Equation (3) there,
  while those of Equation (14) are.}. 
A sample use is prepared with subroutine {\tt ZF\_APV}, with
$N=78$ and $Z=55$ for Caesium.
Flag {\tt TUPV} may be used for a study of the theoretical
uncertainty, the default value is  {\tt TUPV}=1.


\eqnzero
\section{Higher order electroweak corrections to Standard Model observables }
\label{changes}

In {\tt ZFITTER} version 6.42, all  known two- and three loop corrections of
the Standard Model observables $\mwl$, $\seffsf{}$ and $\gz$ have
consistently  been included.  The  corresponding changes began with 
version 6.33, with subsequent additions until version 6.42. 
They are described to  some detail in this section.

\subsection{The $W$ boson mass}
\label{deltaR}

{\tt ZFITTER}, version 6.36  (21 July 2001), contained only  the leading and
next-to-leading  corrections  to  $\Delta r$,  obtained  through  an
expansion in the heavy top quark mass.  They were applied according to
\cite{Degrassi:1994a0,Degrassi:1995ae,Degrassi:1995mc,
Degrassi:1996mg,Degrassi:1996ZZ,Degrassi:1999jd}
and have been implemented in  the package {\tt m2tcor.f}.  The W boson
mass was then evaluated with the relation:
\bq  \mwl= \frac{\mzl}{\sqrt{2}}\sqrt{1+\sqrt{1-\frac{4  \pi  \alpha }
{\sqrt{2} \mzs \gf \lpar 1-\Delta r \rpar }}}\;,
\label{wmass}
\eq
by an iterative procedure, since $\Delta r$ also depends on $\mwl$.
More accurate  results on the  electroweak corrections to  muon decay,
which successively appeared in
\cite{Freitas:2000gg,Freitas:2000nv,Freitas:2001zs,Freitas:2002ja,
Awramik:2003ee,Awramik:2002wn,Awramik:2002wv,Awramik:2002vu,Awramik:2003rn,
Onishchenko:2002ve},   
were  found  in
terms of  exact one-dimensional integral representations  at the order
${\cal O}(\alpha^2)$.   However, since the computation  of general two
loop integrals is  rather slow, only fitting formulas  as published in
\cite{Freitas:2002ja,Awramik:2003rn}  were   implemented  in  {\tt ZFITTER}.
Furthermore,  since typical resummation  prescriptions for  $\Delta r$
(see \cite{Bardin:1995a2} and references therein) 
are  problematic   once  the  complete  two-loop
contributions are  included, the new result has  been directly applied
to $\mwl$.

With  the  option  ${\tt  AMT4}=5$  the complete  fermionic  two  loop
corrections to the $W$ boson mass \cite{Freitas:2002ja} are used. Here
the fermionic corrections encompass all two-loop contributions with at
least one closed fermion loop. In addition to electroweak corrections,
this  setting  also  includes  the  leading  ${\cal  O}(\alpha  \als)$
\cite{Djouadi:1987gn,Djouadi:1988di,Kniehl:1990yc,Halzen:1991je,
Kniehl:1992gu,Kniehl:1993dx,Djouadi:1994ss}
and next-to-leading ${\cal O}(\alpha\als^2)$
\cite{Avdeev:1994db,Chetyrkin:1995ix,Chetyrkin:1995js}              
QCD
corrections. Contrary  to the settings  ${\tt AMT4}\le 4$,  the ${\cal
O}(\alpha \als^2)$  corrections are incorporated  exactly according to
\cite{Chetyrkin:1995js}. Previously,  only the leading  term in $\mts$
was used.

The  assignment ${\tt  AMT4}=6$, which  is used
from Summer 2004  onwards, enables the calculation  of $\mwl$
including  complete  (fermionic   and  bosonic)  ${\cal  O}(\alpha^2)$
corrections        
\cite{Awramik:2002wn,Awramik:2003rn},        
${\cal O}(\alpha\als)$
\cite{Djouadi:1987gn,Djouadi:1988di,Kniehl:1990yc,Halzen:1991je,
Kniehl:1992gu,Kniehl:1993dx,Djouadi:1994ss}   and   ${\cal   O}(\alpha
\als^2)$ \cite{Chetyrkin:1995js,Chetyrkin:1996cf}  QCD corrections, as
well as  leading three-loop corrections  in an expansion in  $\mts$ of
order   ${\cal    O}(\alpha^3)$   and   ${\cal    O}(\alpha^2   \als)$
\cite{Faisst:2003px}.  These contributions were implemented into a fitting
formula of the following form \cite{Awramik:2003rn}:
\begin{eqnarray}
\label{eq:fitformula}
\mwl &=& \mwl^0 - c_1 \, \mathrm{dH} - c_2 \, \mathrm{dH}^2 
       + c_3 \, \mathrm{dH}^4 + c_4 (\mathrm{dh} - 1)
       - c_5 \, \mathrm{d}\alpha + c_6 \, \mathrm{dt} \\
&& {}  - c_7 \, \mathrm{dt}^2 \nonumber 
       - c_8 \, \mathrm{dH} \, \mathrm{dt} 
       + c_9 \, \mathrm{dh} \, \mathrm{dt} - c_{10} \, \mathrm{d}{\alpha_{\mathrm s}}
       + c_{11} \, \mathrm{dZ} ,
\end{eqnarray}
where
\vspace{-0.5cm}
\begin{eqnarray}
&&{\text{\mathrm{dH}}} = \log\left(\frac{\mhl}{\mathrm{\text{100 \,\, \mbox{GeV}}}} \right), 
\hspace{1.4cm}
{\text{\mathrm{dh}}} = \left(\frac{\mhl}{\mathrm{\text{100 \,\, GeV}}}\right)^2, 
 \nonumber \\ 
\hspace{0.51cm}
&&{\text{\mathrm{dt}}} = \left(\frac{m_t}{\mathrm{\text{174.3 {\,\, GeV}}}}\right)^2 - 1, 
\hspace{1cm}
{\text{\mathrm{dZ}}} = \frac{\mzl}{\mathrm{\text{91.1875 {\,\, \mathrm{GeV}}}}} -1, 
\label{eq:pardef}\\ 
&&{\text{\mathrm{d\alpha}}} = \frac{\mathrm{\text{\Delta\alpha}}}{\mathrm{\text{0.05907}}} - 1, 
\hspace{2.1cm}
{\text{\mathrm{d\alpha_{\mathrm s}}}} = \frac{\mathrm{\text{\alpha_{\mathrm s}}}(\mzl)}{\mathrm{\text{0.119}}} - 1 ,
\nonumber
\end{eqnarray}
and the coefficients $\mwl^0, c_1, \ldots, c_{11}$ take the following values
\begin{equation}
\begin{array}{rclrclrcl}
\mwl^0 &=& 80.3799 {\,\, \mathrm{GeV}}, &\quad  c_1 &=& 0.05429 {\,\, \mathrm{GeV}},  &\quad c_2 &=& 0.008939 {\,\, \mathrm{GeV}} ,\\
c_3 &=& 0.0000890 {\,\, \mathrm{GeV}},  &\quad  c_4 &=& 0.000161 {\,\, \mathrm{GeV}}, &\quad c_5 &=& 1.070 {\,\, \mathrm{GeV}} ,  \\
c_6 &=& 0.5256 {\,\, \mathrm{GeV}},     &\quad  c_7 &=& 0.0678 {\,\, \mathrm{GeV}},  &\quad  c_8 &=& 0.00179 {\,\, \mathrm{GeV}} , \\
c_9 &=& 0.0000659 {\,\, \mathrm{GeV}},  &\quad  c_{10}&=& 0.0737 {\,\, \mathrm{GeV}},&\quad c_{11}&=& 114.9 {\,\, \mathrm{GeV}} . 
\end{array} \label{eq:fitparams}
\end{equation}
With eq.~(\ref{eq:fitformula}) the full result for $\mwl$ is approximated to
better than  0.5~MeV over  the range of  $10 {\,\,  \mathrm{GeV}} \leq
\text{\mhl} \leq 1 {\,\,  \mathrm{TeV}}$ if all other experimental input
values vary within their combined $2 \sigma$ region around the central
values, as being used in eq.~(\ref{eq:pardef}).


For the new options ${\tt AMT4}=5,6$ the algorithm used by {\tt ZFITTER} for
the  estimation  of the  theory  error due to the calculation of
$\mwl$,  based  on applying  different
resummation  procedures (see  eq.~(2.104)  of \cite{Bardin:1999yd})
cannot be used anymore, since the next term in the $\mtl$ expansion, 
which is of order ${\cal O}(\gfs m_t^0 \mzs)$,
is   now  known  exactly.    In  contrast,   for  these   options  the
uncertainties   from  unknown  higher   order  contributions   in  the
calculation     of      $\mwl$     have     been      estimated     in
\cite{Freitas:2002ja,Awramik:2003rn} and  are set here to  $\pm 5$ MeV
and $\pm  4$ MeV, respectively. They  can be simulated  by varying the
flag ${\tt DMWW}$ between $-1$ and  1, which is the only relevant flag
to simulate theoretical uncertainties for $\mwl$ for ${\tt AMT4}=5,6$,
whereas {\tt EXPR}, {\tt IHIGS}, {\tt HIG2}, {\tt SCRE} and {\tt SCAL}
are  ignored.  Furthermore,  the   use  of  ${\tt  AMT4}=5,6$  is  not
compatible with the setting ${\tt IMOMS}>1$.

\subsection{The effective leptonic weak mixing angle}
\label{eff_angle}

The amplitude for the decay of the $Z$  boson into a pair of fermions
is parameterized in {\tt ZFITTER} as: 
\bq
V^{\zb\ff\fbf}_{\mu}\lpar\mzs\rpar = 
\left( \ib  \lpar 2\pi\rpar^4\right) \, \ib\,\sqrt{\sqrt{2}\gf\mzs} 
\sqrt{\rho^{\ff}_{\sss{\zb}}}\tcif\gamma_\mu
\lrbr\lpar 1+\gfd \rpar - 4|\qd|\stws\kappa^{\ff}_{\sss{\zb}}\rrbr.
\label{decayrhokappadef}
\eq 
This formula allows to define the effective weak mixing angle as:  
\bq
\seffsf{\ff} = \Reb\bigl(\kZdf{\ff}\bigr)\siws
,
\label{sw2eff}
\eq
where  the coefficients $\rZf$  and $\kZf$  are called  {\em effective
couplings} of $\zb$-decay. 

Complete fermionic two-loop corrections to the leptonic effective weak
mixing     angle      $\sin^2     \theta^{\rm     lept}_{\rm     eff}$
\cite{Awramik:2004ge,Awramik:2004qv,Awramik:2004rt}
are  included  with  the  option  ${\tt AMT4}=6$.   They  improve  the
realizations  for ${\tt  AMT4}  \le 5$  by  including all  electroweak
two-loop contributions  with at least  one closed fermion loop  and go
beyond the leading
\cite{Barbieri:1992nz,Barbieri:1993dq,Fleischer:1993ub,Fleischer:1995cb}
and next-to-leading \cite{Degrassi:1997ps}  terms in an  expansion in
$\mts$.   The corrections  have  been implemented  directly using  the
numerical  fitting  formula  as  published  in  \cite{Awramik:2004ge},
thereby assuring a fast evaluation:
\begin{eqnarray}
\label{formula}
{\sin^2\theta^{\mbox{\footnotesize lept}}_{\mbox{\footnotesize eff}}} 
&=& s_0 + d_1 L_H + d_2  L_H^2 + d_3  L_H^4 + d_4  (\Delta_H^2 -1) + d_5  \Delta_\alpha 
\nonumber \\ && 
+ d_6  \Delta_t + d_7  \Delta_t^2 + d_8  \Delta_t  (\Delta_H -1)  + d_9  \Delta_{\alpha_s} + d_{10} \Delta_Z,
\end{eqnarray}
with
\begin{eqnarray}
\vspace{0.2cm}
&&L_H = \log\left(\frac{\mhl}{\text{\mathrm{100 \mbox{ GeV}}}}\right),
\hspace{1cm}
\Delta_H = \frac{\mhl}{\text{\mathrm{100 \mbox{ GeV}}}}, 
\nonumber \\ 
\vspace{0.2cm}
&&\Delta_\alpha = \frac{\text{\mathrm{\Delta \alpha}}}{\text{\mathrm{0.05907}}}-1,
\hspace{1.8cm}
\Delta_t = \left(\frac{m_t}{\text{\mathrm{178.0 \mbox{ GeV}}}}\right)^2 -1, 
\\ 
\vspace{0.2cm}
&&\Delta_{\alpha_s} = \frac{\text{\mathrm{\alpha_s}}(\mzl)}{\text{\mathrm{0.117}}}-1,
\hspace{1.7cm}
\Delta_Z = \frac{\mzl}{\text{\mathrm{91.1876 \mbox{ GeV}}}} -1, 
\nonumber
\end{eqnarray}
and
\begin{equation}
\begin{array}{lll}
s_0 = 0.2312527, &\quad d_1 =  4.729 \times 10^{-4}, &\quad d_2 = 2.07
\times  10^{-5}, \\  d_3 =  3.85 \times  10^{-6}, &\quad  d_4  = -1.85
\times 10^{-6}, &\quad d_5 = 0.0207,  \\ d_6 = -0.002851, &\quad d_7 =
1.82 \times 10^{-4}, &\quad d_8 =  -9.74 \times 10^{-6}, \\ d_9 = 3.98
\times 10^{-4}, &\quad d_{10} = -0.655.
\end{array}
\end{equation}
In  addition to the  electroweak one-  and two-loop  corrections, this
formula    also   includes    two-loop    ${\cal   O}(\alpha    \als)$
\cite{Djouadi:1987gn,Djouadi:1988di,Kniehl:1990yc,Halzen:1991je,
Kniehl:1992gu,Kniehl:1993dx,Djouadi:1994ss}   and   three-loop  ${\cal
O}(\alpha    \als^2)$   \cite{Chetyrkin:1995js,Chetyrkin:1996cf}   QCD
corrections,  as well  as  leading  three-loop  corrections in  an
expansion  in   $\mts$  of  order  ${\cal   O}(\alpha^3)$  and  ${\cal
O}(\alpha^2 \als)$  \cite{Faisst:2003px}. In contrast  to the settings
${\tt  AMT4}<6$,  the  three-loop  QCD  corrections  are  incorporated
exactly  according to \cite{Chetyrkin:1995js},  which goes  beyond the
leading  term  in  $\mts$  used previously.  Equation  (\ref{formula})
reproduces the  exact calculation with maximal  and average deviations
of $4.5\times10^{-6}$  and $1.2\times 10^{-6}$,  respectively, as long
as the  input parameters  stay within their  $2\sigma$ ranges  and the
Higgs boson mass in the range 10 GeV $\leq \mhl \leq$ 1 TeV.

With  the new  option ${\tt  AMT4}=6$ the  theory error  in the
calculation of $\sin^2 \theta^{\rm lept}_{\rm eff}$ is  no longer
obtained  through different  resummation formulas  (see  eq.~(2.134) of
\cite{Bardin:1999yd}).   Instead,  the  value  of  $  \pm  4.9  \times
10^{-5}$, as  estimated in  \cite{Awramik:2004ge}, is used.   For this
case the theory uncertainty can be simulated by varying the flag ${\tt
DSWW}$ between $-1$ and 1.

\subsection{Partial and total $Z$  boson widths}

The $Z$   boson decay width  is defined through the  effective couplings,
$\kZdf{f}$ and $\rZf$, of eq.~(\ref{decayrhokappadef}).
The form factor $\kZdf{l}$ is  obtained from the effective weak mixing
angle  $\sin^2  \theta^{\rm  lept}_{\rm  eff}$, as  presented  in 
subsection \ref{eff_angle}.  For $Z$  boson decays into a pair of
fermions  $f$, $f  \ne  l$,  the two-loop  corrections  to $\kZf$  are
currently implemented in an  approximate way only, while for $\Gamma(Z
\rightarrow  b\bar{b})$, no  two-loop corrections  beyond  the leading
$\mtq$ term are available.
For  the  form factor  $\rZf$,  the  two-loop  corrections beyond  the
next-to-leading $\mts$  expansion are still missing.  This quantity is
therefore computed identically for the choices ${\tt AMT4}=4,5,6$.

\subsection{Electroweak form factors for $\fep\fem\to\ff\fbf$}

The radiative corrections to the cross-sections and asymmetries for 
the process $\fep\fem\to\ff\fbf$ are parameterized in {\tt ZFITTER} 
by electroweak form factors,  
$\rho_{ef},\kappa_{e},\kappa_{f}$ and $\kappa_{ef}$,  
defined through a $Z$-boson exchange amplitude: 
{\small{
\bqa
{\cal A}
_{\sss{\zb}}(\sman,\tman)&=& 
        \ib\,e^2\,4\,\tcie\tcif\frac{\chi_{\sss{Z}}(\sman)}{\sman} 
        \rho_{ef}(\sman,\tman)
        \biggl\{
        \gadu{\mu}{\lpar 1+\gfd \rpar }
        \otimes \gadu{\mu} { \lpar 1+\gfd \rpar}    
\nll\nll &&
-4 |\qe | \stws \kappa_e(\sman,\tman)
        \gadu{\mu} \otimes \gadu{\mu}{\lpar1+\gfd\rpar}
       -4 |\qf | \stws \kappa_f(\sman,\tman)
        \gadu{\mu} {\lpar 1+\gfd \rpar } 
        \otimes \gadu{\mu}                              
\nll\nll &&
+16 |\qe \qf| \stwf \kappa_{ef}(\sman,\tman)
        \gadu{\mu} \otimes \gadu{\mu} \biggr\}, 
\label{processrhokappadef}
\eqa
}}
where 
\bq
\chi_Z(\sman) =
\frac{\gf}{\sqrt{2}}\frac{\mzs}{8\pi\alpha}\frac{\sman}{\sman-m^2_{\sss{Z}}}
\text{\mathrm{\hspace{1cm}  and  \hspace{1cm}}}
m^2_{\sss{Z}} = \mzs - \ib\mzl\gz(\sman).
\label{mZ2}
\eq

In the leading pole approximation, the following relations hold:
\bq
\rho_{ef} = \sqrt{\rho^{e}_{\sss{\zb}} \,\rho^{\ff}_{\sss{\zb}}}, \;\;\;\;\;
\kappa_{e} = \kappa^{e}_{\sss{\zb}}, \;\;\; \;\;
\kappa_{f} = \kappa^{f}_{\sss{\zb}}, \;\; \; \;\;
\kappa_{ef} = \kappa^e_{\sss{\zb}} \kappa^f_{\sss{\zb}}.
\eq
However, generally,
$\rho_{ef}$, $\kappa_{e}$, $\kappa_{f}$, and $\kappa_{ef}$ also include
$\gamma$-$Z$ interference effects, corrections from non-resonant $Z$
and $\gamma$ exchanges, and non-factorizable box contributions.
With the option ${\tt AMT4}=6$ the coefficients  $\kappa_{e}$ 
and $\kappa_{f}$, for $f \ne b$,  are calculated with
all known two- and three-loop effects, as discussed in 
the previous sections. 
The corresponding factorizable vertex corrections to $\kappa_{ef}$ are
incorporated similarly. The electroweak box contributions to the form 
factors include the complete
one-loop order correction, which is sufficient for present precision.
For the form factor $\rho_{ef}$, complete two-loop corrections 
are still missing, and it is computed including 
two-loop corrections of the order ${\cal O}(\gfs\mts\mzs)$ only  
for all options ${\tt AMT4}=4,5,6$.

For the $Z \to  b\bar{b}$ channel, no two-loop electroweak corrections
beyond  the  leading $\mtq$  terms  are  available  so far. Furthermore,
in  {\tt ZFITTER} versions before 6.42, a mismatch occurred in 
the treatment of
the $b\bar{b}$ final  state. The subroutine {\tt GDEGNL}, which computes the
leading two-loop corrections,  is called at
two  places   in  the  program: 
({\it i})  in  subroutine  {\tt   ZWRATE} 
the effective couplings introduced in section \ref{eff_angle}
are used for the  computation of $Z$ partial widths; 
({\it ii}) the interfaces {\tt ZUTHSM}, {\tt  ZUTPSM}, {\tt  ZULRSM} 
and {\tt  ZUATSM}, on  the other hand,  calculate  the cross-sections  
and  asymmetries  from the  weak form factors 
$\rho_{ef}(s,t)$, $\kappa_e(s,t)$, $\kappa_f(s,t)$
and $\kappa_{ef}(s,t)$ in the subroutine {\tt ROKANC}.
For ${\tt  INDF}=9$, i.e.  the $b\bar{b}$ final state, all  these form
factors  were calculated  in one-loop  approximation with  the leading
$\mtq$ two-loop term.  As a result, in {\tt  ROKANC} the initial-state
$Ze^+e^-$  form  factors for  all  other  final  states are  generated
including next-to-leading 
two-loop  corrections, while for the  $b\bar{b}$ final state
these corrections were missing.
This mismatch  also affects the {\tt ZFITTER}  interfaces {\tt ZUXSA},
{\tt  ZUTAU} and  {\tt ZUXSA2},  which use  the language  of effective
couplings
\cite{Bardin:1999yd}, since they are defined to coincide 
as close as possible with
the  complete  Standard  Model  prediction  in  {\tt  ROKANC}  if  the
effective couplings coincide with their Standard Model analogue.

The problem has  been alleviated since {\tt ZFITTER}  version 6.41 (15
October 2004). In contrast to the older implementations, $\kappa_e(s,t)$
and $\kappa_f(s,t)$  are not  treated symmetrically anymore  for ${\tt
INDF}=9$,  but  two-loop   electroweak  corrections  are  included  in
$\kappa_e(s,t)$    for   ${\tt    AMT4}   \geq    4$,   not    yet   in
$\kappa_b(s,t)$.    The     treatment    of    $\rho_{ef}(s,t)$    and
$\kappa_{ef}(s,t)$ has been changed  
accordingly\footnote{In {\tt ZFITTER} version 6.41 an error occured 
in the treatment of a form factor $\rho_{ef}(s,t)$. It affected 
the $b\bar{b}$ cross-section only, and was corrected in 
{\tt ZFITTER} version 6.42.}. 
Here one can use the
fact that  the presently  known two-loop contributions  factorize into
initial-state  and  final-state   corrections.  For  a  more  detailed
discussion see \cite{Freitas:2004mn}.

\eqnzero
\section{Input parameters\label{input} and pseudo observables}

For the calculations of pseudo observables (POs), such as partial $Z$ decay
widths or effective coupling constants, \zf\ uses the {\tt
DIZET} package which calculates such quantities employing the 
{\em on-mass-shell (OMS)} renormalization scheme
\cite{Bardin:1980fet,Bardin:1982svt} within the Standard Model.
The user has to provide a set of values for the so-called input 
parameter set,
which are then used by \zf\ and {\tt DIZET} to calculate the POs.  A
standard set of imput parameters is given by:
\begin{itemize}
\item The electromagnetic coupling constant at the $Z$ pole, or, more 
precisely, its shift due to the  5-quark flavour hadronic vacuum polarisation, 
$\Delta\alpha^{(5)}_{\mathrm{had}}$ at the $Z$ pole;
\item The strong coupling constant at the $Z$ pole, $\alpha_S$;
\item The pole masses of $Z$ boson, top quark and Higgs boson.
\end{itemize}
These variables are the main physics values to be provided to \zf\ and
{\tt DIZET} by the user. The Fermi constant is treated as a constant
in the program (see flag {\tt GFER}). In order to use the
user-supplied value of the hadronic vacuum polarisation, flag {\tt ALEM=2}
must be used.  In order to use the latest set of electroweak
radiative corrections (see~\cite{Awramik:2003rn,Awramik:2004ge} 
and Section~\ref{changes}), flag {\tt AMT4=6} must be used.


Selected POs calculated by \zf\ and {\tt DIZET} are listed in
\tabn{tabPOs}, namely: (i) the $\wb$ boson mass and the on-shell
electroweak mixing angle $\siws=1-\mws/\mzs$; (ii) the partial $Z$ decay
widths, including the invisible width, simply equal to $3\gn$, and the
the total hadronic width, equal to the sum of the five-flavour
quarkonic widths, and the total width; (iii) ratios of widths and pole
cross-sections as used by, e.g., the LEP EWWG; (iv) the effective
electroweak mixing angle $\seffsf{\ff}$ according to \eqn{sw2eff} and
the $\rho_f$ parameter $\rho_f=\mathrm{Re}(\rho^f_Z)$, for leptons and
heavy fermions $b$ and $c$; and (v) the asymmetry parameter ${\cal
A}_{\ff}$ and the forward-backward pole asymmetries $\afba{0 f}$ for
leptons and heavy fermions $b$ and $c$.

The partial decay widths are defined inclusively, i.e., they contain
all real and virtual corrections. The ratios for leptons $l$ and
quarks $q$ are defined as:
\bq
R^0_{\fl}~=~\frac{\gh}{\gll} \qquad\qquad R^0_{\fq}~=~\frac{\gqq}{\gh}\,,
\eq
while the hadronic and leptonic pole cross-sections
are defined as:
\bq
\sigma^0_{\had} ~=~ 12\pi\,\frac{\gel\gh}{\mzs\gzs} \qquad \qquad
\sigma^0_{l} ~=~ 12\pi\,\frac{\gel\gl}{\mzs\gzs} \,.
\eq

The complex variable $g^f_Z$ is defined as the ratio:
\bq
\Rvaz{\ff} 
~=~ \frac{\vc{\ff}}{\ac{\ff}}
~=~ 1-4|Q_f| \kZdf{\ff}\siws
\label{varatiorez1}
\eq
of the complex effective vector and axial couplings of the $Z$ boson to 
the fermion $f$:
\ba
\vc{\ff} \, &~=~&  \, \sqrt{\rho^f_Z}\,(T^f_3  - 2 Q_f \kZdf{\ff}\siws) \,,
\\
\ac{\ff} \, &~=~&  \, \sqrt{\rho^f_Z}\,T^f_3.
\label{ratio}
\ea
Hence one may write: 
\bq
\text{\mathrm{Re}}(\Rvaz{\ff}) 
~=~1-4|Q_f| \sin^2\theta^f_{\mathrm{eff}}.
\label{varatiorez2}
\eq
The complex form factors $\rho^f_Z$, $\kappa^f_Z$ are defined through
the amplitude of the $Z$ boson decay into a pair of fermions, as in
eq.~(\ref{decayrhokappadef}).

With these definitions, the asymmetry parameters are then given
in terms of the real part of $g^f_Z$:
\bq
{\cal A}_{\ff}   ~=~
2\frac{\Reb\,\Rvaz{\ff}}{1+\Bigl(\Reb\,\Rvaz{\ff}\Bigr)^2}\,.
\label{def_coupling}
\eq
The forward-backward pole asymmetries are mere combinations of these
asymmetry parameters:
\bq
\afba{0 f} ~=~\frac{3}{4}\,{\cal A}_{\fe}\,{\cal A}_{\ff}.
\label{def_afbs}
\eq

In terms of the real part of the complex $\rho^f_Z$ parameter,
$\rho_f=\mathrm{Re}(\rho^f_Z)$, and the (real) effective electroweak
mixing angle $\seffsf{\ff}$ defined earlier, the real effective vector
and axial-vector coupling constants as quoted by the LEP EWWG are then
defined as:
\ba 
g_{Af} & ~=~ & \sqrt{\rho_f}\, T^f_3 \\
g_{Vf} & ~=~ & \sqrt{\rho_f}\,(T^f_3-2Q_f\sin^2\theta^f_{\mathrm{eff}})\,.  
\ea 
With these definitions, it follows that:
\bq
\frac{g_{Vf}}{g_{Af}} ~=~ \text{\mathrm{Re}}(\Rvaz{\ff}) \,,
\eq
so that the asymmetry parameters
are then equivalently given as:
\bq
{\cal A}_{\ff} ~ = ~ 2\frac{g_{Vf}/g_{Af}}{1+(g_{Vf}/g_{Af})^2} \,.
\eq



\begin{table}[tb]
\vspace*{1cm}
\renewcommand{\arraystretch}{1.2}
\begin{center}
\begin{tabular}[t]{|c||r|}
\hline
Observable             &  Value     \\
\hline
\hline
$\mwl\,$[GeV]          &  80.3613   \\
$\siws$                &  0.22335   \\
\hline
\hline
$\gn\,$[MeV]           &  167.219   \\ 
$\gel\,$[MeV]          &   83.990   \\
$\gmu\,$[MeV]          &   83.990   \\
$\gt\,$[MeV]           &   83.800   \\
$\gc\,$[MeV]           &  299.966   \\ 
$\gbq\,$[MeV]          &  375.729   \\  
\hline
$\gi\,$[GeV]           &  0.501657  \\
$\gh\,$[GeV]           &  1.741507  \\
$\gz\,$[GeV]           &  2.494944  \\ 
\hline
\end{tabular}
\ 
\begin{tabular}[t]{|c||r|}
\hline
Observable             &  Value     \\
\hline
\hline
$R^0_{\fl}$            &  20.73462  \\
$R^0_{\fb}$            &  0.215750  \\
$R^0_{\fc}$            &  0.172245  \\
\hline
$\sigma^0_{\had}\,$[nb]&  41.4826   \\
$\sigma^0_{\lep}\,$[nb]&   2.0006   \\
\hline
\hline
$\seffsf{\rm{lept}}$   &  0.231548  \\
$\seffsf{\fb}$         &  0.233032  \\ 
$\seffsf{\fc}$         &  0.231442  \\
\hline
$\rhoe$                &  1.005165  \\
$\rhoi{\fb}$           &  0.993943  \\
$\rhoi{\fc}$           &  1.005860  \\
\hline
\end{tabular}
\ 
\begin{tabular}[t]{|c||r|}
\hline
Observable             &  Value     \\
\hline
\hline
${\cal A}_{\fe}$       &  0.146813  \\  
${\cal A}_{\fb}$       &  0.934554  \\ 
${\cal A}_{\fc}$       &  0.667779  \\
\hline
$\afba{0,\fl}$         &  0.016166  \\
$\afba{0,\fb}$         &  0.102904  \\
$\afba{0,\fc}$         &  0.073529  \\
\hline
\end{tabular}
\end{center} \ 
\caption[Predictions for pseudo observables from {\tt ZFITTER}]
{\it
Predictions for pseudo observables calculated with \zf\ and {\tt DIZET}
for $\Delta\alpha^{(5)}_{\mathrm{had}}=0.02758$, $\alpha_S=0.118$,
$M_Z=91.1875~GeV$, $M_t=175~GeV$ and $M_H=150~GeV$, and  flags
{\tt ALEM=2} and {\tt AMT4=6}.
\label{tabPOs}}
\end{table}

\subsection{Pseudo observables in common blocks of {\tt DIZET} \label{POCOMM}} 
The channel dependent quantities are stored in common block
{\tt COMMON/CDZRKZ/} and in array {\tt PARTZ(0:11)}:
\begin{small}
\begin{verbatim}

 COMMON/CDZRKZ/ARROFZ(0:10),ARKAFZ(0:10),ARVEFZ(0:10),ARSEFZ(0:10)
&             ,AROTFZ(0:10),AIROFZ(0:10),AIKAFZ(0:10),AIVEFZ(0:10)
*---
 DIMENSION NPAR(25),ZPAR(30),PARTZ(0:11),PARTW(3)
*---
 SUBROUTINE DIZET(NPAR,AMW   !  NPAR : FLAGS; AMW : INPUT/OUTPUT
&                     ,AMZ,AMT,AMH,DAL5H,V_TBA,ALSTR   !  INPUT
&                     ,ALQED,ALSTRT,ZPAR,PARTZ,PARTW)  !  OUTPUT

\end{verbatim}
\end{small}
\noindent The correspondences have not been changed compared to
Section 2.5.1 of \cite{Bardin:1999yd}:
\bqa
\mbox{\tt ARROFZ(0:10)    }&=&(\rZdf{\ff})^{'}\,,
\\
\mbox{\tt AROTFZ(0:10)    }&=&\Reb\,\rZdf{\ff}\,,
\\
\mbox{\tt ARKAFZ(0:10)    }&=&\Reb\,\kZdf{\ff}\,,
\\
\mbox{\tt ARVEFZ(0:10)    }&=&\Reb\,\Rvaz{\ff}\,,
\\
\mbox{\tt AIROFZ(0:10)    }&=&\Imb\,\rZdf{\ff}\,,
\\
\mbox{\tt AIKAFZ(0:10)    }&=&\Imb\,\kZdf{\ff}\,,
\\
\mbox{\tt AIVEFZ(0:10)    }&=&\Imb\,\Rvaz{\ff}\,,
\\
\mbox{\tt ARSEFZ(0:10)    }&=&\seffsf{\ff}\,,
\\
\quad\mbox{\tt PARTZ(0:11)}&=&\gff\,,
\label{CDZRKZ}
\eqa
The usual {\tt ZFITTER} channel assignments as given in 
Figure 5 of \cite{Bardin:1999yd} are used,
and all quantities but $(\rZdf{\ff})^{'}$ have been introduced already.
The  $(\rZdf{\ff})^{'}$ is discussed in Section 2.5.1 of \cite{Bardin:1999yd}.
We mention also here that
both options {\tt MISC=1,0} (with $(\rZdf{\ff})^{'}$ or with $|\rZdf{\ff}|$) 
are used in the {\em Model Independent interfaces} of {\tt ZFITTER},
see \appendx{interfaces}.
%
%
%

Note that since the convention of the LEP EWWG is to use the real
parameter $\rho_f=\mathrm{Re}(\rho^f_Z)$, not $(\rho^f_Z)'$, the array
{\tt AROTFZ} is generally used, while {\tt ARROFZ, RENFAC and SRENFC}
are usually ignored.



\eqnzero
\section{Subroutine {\tt ZFTEST}\label{zftest} }
The \zf\ distribution includes subroutine {\tt ZFTEST}.
With  {\tt ZFTEST} the user may test whether \zf\ has been properly
installed.
The subroutine calculates cross-sections and asymmetries as
functions of \RS\ near the \Z\ peak, below, and above.
A sample file {\tt zfmai6$\_$42.f} runs  {\tt ZFTEST}:
\begin{verbatim}
*
* MAIN to call ZFTEST with version 6.42
*
      CALL ZFTEST(0)
      END
\end{verbatim}
The numerical output should reproduce the Tables given in \subsect{results}.
\subsection{{\tt ZFTEST} results\label{results}}
This Section contains the standard test-outputs, 
produced by a call to {\tt ZFTEST(0)}.
The argument of {\tt ZFTEST} sets the flag {\tt MISC}.
The default value is {\tt MISC}=0.
The user's result should be identical to the sample output, apart from
a possible flip in the last digit.

\vfill

\pagebreak
\begin{turn}{90}
\begin{minipage}{\textheight}

\begin{Verbatim}[fontsize=\tiny]
 ******************************************************
 ******************************************************
 **           This is ZFITTER version 6.42           **
 **                   05/05/18                       **
 ******************************************************
 ** http://www.ifh.de/theory/publist.html            **
 ******************************************************

 ZUINIT> ZFITTER defaults:

 ZFITTER flag values:
 AFBC: 1 SCAL: 0 SCRE: 0 AMT4: 4 BORN: 0
 BOXD: 1 CONV: 1 FINR: 1 FOT2: 3 GAMS: 1
 DIAG: 1 INTF: 1 BARB: 2 PART: 0 POWR: 1
 PRNT: 0 ALEM: 3 QCDC: 3 VPOL: 1 WEAK: 1
 FTJR: 1 EXPR: 0 EXPF: 0 HIGS: 0 AFMT: 3
 CZAK: 1 PREC:10 HIG2: 0 ALE2: 3 GFER: 2
 ISPP: 2 FSRS: 1 MISC: 0 MISD: 1 IPFC: 5
 IPSC: 0 IPTO: 3 FBHO: 0 FSPP: 0 FUNA: 0
 ASCR: 1 SFSR: 1 ENUE: 1 TUPV: 1 DMWW: 0
 DSWW: 0


 ZFITTER cut values:
   INDF  ICUT    ACOL    EMIN    S_PR    ANG0    ANG1     SPP
      0    -1    0.00    0.00    0.00    0.00  180.00    0.00
      1    -1    0.00    0.00    0.00    0.00  180.00    0.00
      2    -1    0.00    0.00    0.00    0.00  180.00    0.00
      3    -1    0.00    0.00    0.00    0.00  180.00    0.00
      4    -1    0.00    0.00    0.00    0.00  180.00    0.00
      5    -1    0.00    0.00    0.00    0.00  180.00    0.00
      6    -1    0.00    0.00    0.00    0.00  180.00    0.00
      7    -1    0.00    0.00    0.00    0.00  180.00    0.00
      8    -1    0.00    0.00    0.00    0.00  180.00    0.00
      9    -1    0.00    0.00    0.00    0.00  180.00    0.00
     10    -1    0.00    0.00    0.00    0.00  180.00    0.00
     11    -1    0.00    0.00    0.00    0.00  180.00    0.00



 ZFITTER input parameters:
 DAL5H =  0.0280398093
 ALQED5=  128.886183
  ZMASS =   91.18760;  TMASS =  178.00000
  HMASS =  100.00000
  DAL5H =    0.02804; ALQED5 =  128.88618
  ALFAS =    0.11700;  ALFAT =    0.10637

 ZFITTER intermediate results:
  WMASS =   80.39522; SIN2TW =    0.22270

 ALPHST =    0.11700;
 QCDCOR =  1.00000 1.03938 1.04639 1.03877 1.03177 1.03941 1.04611 1.03877 1.03177 1.03938 1.04639 1.03937 1.02506-0.00002 0.19863

 CHANNEL         WIDTH         RHO_F_R        RHO_F_T        SIN2_EFF
 -------        -------       --------       --------       --------
 nu,nubar       167.299       1.008546       1.008546       0.231036
 e+,e-           84.037       1.005790       1.005643       0.231417
 mu+,mu-         84.037       1.005790       1.005643       0.231417
 tau+,tau-       83.847       1.005790       1.005643       0.231417
 u,ubar         300.128       1.006390       1.006339       0.231310
 d,dbar         383.033       1.007315       1.007306       0.231183
 c,cbar         300.059       1.006390       1.006339       0.231310
 s,sbar         383.033       1.007315       1.007306       0.231183
 t,tbar           0.000       0.000000       0.000000       0.000000
 b,bbar         375.593       0.993885       0.993885       0.233004
 hadron        1741.846
 total         2495.664
\end{Verbatim}
\end{minipage}
\end{turn}
\pagebreak
\begin{turn}{90}
\begin{minipage}{\textheight}

\begin{Verbatim}[fontsize=\tiny]

 W-widths
 lept,nubar     679.896
 down,ubar     1412.691
 total         2092.587

 C1U(D)2U(D)=-.1889062284  0.3412800711  -.0371905250  0.0234933913

 USE OF AN OBSOLETED OPTION, ICUT=0;
 Might be useful for backcompatibility with 5.xx;
 Presently, ICUT=2,3 recommended for realistic cuts.
 ZFITTER flag values:
 AFBC: 1 SCAL: 0 SCRE: 0 AMT4: 6 BORN: 0
 BOXD: 1 CONV: 1 FINR: 1 FOT2: 3 GAMS: 1
 DIAG: 1 INTF: 1 BARB: 2 PART: 0 POWR: 1
 PRNT: 0 ALEM: 3 QCDC: 3 VPOL: 1 WEAK: 1
 FTJR: 1 EXPR: 0 EXPF: 0 HIGS: 0 AFMT: 3
 CZAK: 1 PREC:10 HIG2: 0 ALE2: 3 GFER: 2
 ISPP: 1 FSRS: 1 MISC: 0 MISD: 1 IPFC: 5
 IPSC: 0 IPTO: 3 FBHO: 0 FSPP: 0 FUNA: 0
 ASCR: 1 SFSR: 1 ENUE: 1 TUPV: 1 DMWW: 0
 DSWW: 0


  SQRT(S) =   35.

      <-------------- Cross Section -------------->  <------- Asymmetry ------->  <--Tau_Pol-->  <----A_LR--->
 INDF   ZUTHSM   ZUXSEC    ZUXSA   ZUXSA2   ZUXAFB   ZUTHSM  ZUXSA ZUXSA2 ZUXAFB ZUTPSM  ZUTAU  ZULRSM  ZUALR
    0  0.00057  0.00057
    1  0.09502  0.09502  0.09502  0.09502  0.09502  -0.0647-0.0647-0.0647-0.0647
    2  0.09527  0.09527  0.09527  0.09527  0.09527  -0.0646-0.0646-0.0646-0.0646
    3  0.09410  0.09410  0.09410  0.09410  0.09410  -0.0644-0.0644-0.0644-0.0644 0.0068 0.0068
    4  0.13269  0.13269  0.13269                    -0.1109-0.1109                            -0.0562 0.0000
    5  0.03386  0.03386  0.03386                    -0.1995-0.1995                            -0.1951 2.1932
    6  0.13278  0.13278  0.13278                    -0.1113-0.1113                            -0.0562 2.1932
    7  0.03386  0.03386  0.03386                    -0.1995-0.1995                            -0.1951 2.0196
    8  0.00000  0.00000  0.00000                     0.0000 0.0000                             0.0000 2.1951
    9  0.03236  0.03236  0.03236                    -0.2112-0.2112                            -0.2028 0.0000
   10  0.36555  0.36555                                                                       -0.0949 0.0000
   11  1.47314  1.46905  1.46905  1.46905            0.8513 0.8491 0.8491



  SQRT(S) =   65.

      <-------------- Cross Section -------------->  <------- Asymmetry ------->  <--Tau_Pol-->  <----A_LR--->
 INDF   ZUTHSM   ZUXSEC    ZUXSA   ZUXSA2   ZUXAFB   ZUTHSM  ZUXSA ZUXSA2 ZUXAFB ZUTPSM  ZUTAU  ZULRSM  ZUALR
    0  0.00558  0.00558
    1  0.03188  0.03188  0.03188  0.03188  0.03188  -0.3671-0.3671-0.3671-0.3671
    2  0.03196  0.03196  0.03196  0.03196  0.03196  -0.3663-0.3663-0.3663-0.3663
    3  0.03196  0.03196  0.03196  0.03196  0.03196  -0.3651-0.3651-0.3651-0.3651 0.0250 0.0250
    4  0.04973  0.04973  0.04973                    -0.4738-0.4738                            -0.2291 0.0000
    5  0.02195  0.02195  0.02195                    -0.4876-0.4876                            -0.4444 2.1932
    6  0.04975  0.04975  0.04975                    -0.4746-0.4746                            -0.2291 2.1932
    7  0.02195  0.02195  0.02195                    -0.4876-0.4876                            -0.4444 2.0196
    8  0.00000  0.00000  0.00000                     0.0000 0.0000                             0.0000 2.1951
    9  0.02165  0.02165  0.02165                    -0.4980-0.4980                            -0.4500 0.0000
   10  0.16503  0.16503                                                                       -0.3153 0.0000
   11  0.45484  0.45474  0.45473  0.45473            0.8370 0.8366 0.8366
\end{Verbatim}
\end{minipage}
\end{turn}
\pagebreak
\begin{turn}{90}
\begin{minipage}{\textheight}

\begin{Verbatim}[fontsize=\tiny]


  SQRT(S) =   89.1875992

      <-------------- Cross Section -------------->  <------- Asymmetry ------->  <--Tau_Pol-->  <----A_LR--->
 INDF   ZUTHSM   ZUXSEC    ZUXSA   ZUXSA2   ZUXAFB   ZUTHSM  ZUXSA ZUXSA2 ZUXAFB ZUTPSM  ZUTAU  ZULRSM  ZUALR
    0  0.79464  0.79464
    1  0.41291  0.41291  0.41291  0.41291  0.41291  -0.1885-0.1885-0.1885-0.1885
    2  0.41309  0.41309  0.41309  0.41309  0.41309  -0.1885-0.1885-0.1885-0.1885
    3  0.41223  0.41223  0.41223  0.41223  0.41223  -0.1888-0.1888-0.1888-0.1888-0.1206-0.1206
    4  1.44131  1.44131  1.44131                    -0.0473-0.0473                             0.0841 0.0000
    5  1.81653  1.81653  1.81653                     0.0544 0.0544                             0.1031 2.1932
    6  1.44105  1.44105  1.44105                    -0.0474-0.0474                             0.0841 2.1932
    7  1.81654  1.81654  1.81653                     0.0544 0.0544                             0.1031 2.0196
    8  0.00000  0.00000  0.00000                     0.0000 0.0000                             0.0000 2.1951
    9  1.78022  1.78022  1.78022                     0.0542 0.0542                             0.1025 0.0000
   10  8.29565  8.29565                                                                        0.0963 0.0000
   11  0.70082  0.70101  0.70090  0.70090            0.4313 0.4312 0.4312


  SQRT(S) =   91.1875992

      <-------------- Cross Section -------------->  <------- Asymmetry ------->  <--Tau_Pol-->  <----A_LR--->
 INDF   ZUTHSM   ZUXSEC    ZUXSA   ZUXSA2   ZUXAFB   ZUTHSM  ZUXSA ZUXSA2 ZUXAFB ZUTPSM  ZUTAU  ZULRSM  ZUALR
    0  2.91319  2.91319
    1  1.47692  1.47692  1.47692  1.47692  1.47692   0.0003 0.0003 0.0003 0.0003
    2  1.47745  1.47745  1.47745  1.47745  1.47745   0.0003 0.0003 0.0003 0.0003
    3  1.47438  1.47438  1.47438  1.47438  1.47438   0.0001 0.0001 0.0001 0.0001-0.1431-0.1431
    4  5.24201  5.24201  5.24201                     0.0615 0.0615                             0.1418 0.0000
    5  6.67109  6.67109  6.67109                     0.0971 0.0971                             0.1441 2.1932
    6  5.24108  5.24108  5.24109                     0.0616 0.0616                             0.1418 2.1932
    7  6.67112  6.67112  6.67113                     0.0971 0.0971                             0.1441 2.0196
    8  0.00000  0.00000  0.00000                     0.0000 0.0000                             0.0000 2.1951
    9  6.54153  6.54153  6.54153                     0.0982 0.0982                             0.1444 0.0000
   10 30.36683 30.36683                                                                        0.1434 0.0000
   11  1.37174  1.37182  1.37174  1.37174            0.1869 0.1869 0.1869


  SQRT(S) =   93.1875992

      <-------------- Cross Section -------------->  <------- Asymmetry ------->  <--Tau_Pol-->  <----A_LR--->
 INDF   ZUTHSM   ZUXSEC    ZUXSA   ZUXSA2   ZUXAFB   ZUTHSM  ZUXSA ZUXSA2 ZUXAFB ZUTPSM  ZUTAU  ZULRSM  ZUALR
    0  1.22102  1.22102
    1  0.62880  0.62880  0.62880  0.62880  0.62880   0.1209 0.1209 0.1209 0.1209
    2  0.62905  0.62905  0.62905  0.62905  0.62905   0.1209 0.1209 0.1209 0.1209
    3  0.62781  0.62781  0.62781  0.62781  0.62781   0.1208 0.1208 0.1208 0.1208-0.1530-0.1530
    4  2.21270  2.21270  2.21270                     0.1293 0.1293                             0.1763 0.0000
    5  2.80592  2.80592  2.80592                     0.1235 0.1235                             0.1695 2.1932
    6  2.21234  2.21234  2.21235                     0.1295 0.1295                             0.1763 2.1932
    7  2.80593  2.80593  2.80594                     0.1235 0.1235                             0.1695 2.0196
    8  0.00000  0.00000  0.00000                     0.0000 0.0000                             0.0000 2.1951
    9  2.75285  2.75285  2.75285                     0.1253 0.1253                             0.1704 0.0000
   10 12.78974 12.78953                                                                        0.1720 0.0000
   11  0.52345  0.52335  0.52336  0.52336            0.2174 0.2174 0.2174

\end{Verbatim}
\end{minipage}
\end{turn}

\pagebreak

\begin{turn}{90}
\begin{minipage}{\textheight}

\begin{Verbatim}[fontsize=\tiny]

  SQRT(S) =   100.

      <-------------- Cross Section -------------->  <------- Asymmetry ------->  <--Tau_Pol-->  <----A_LR--->
 INDF   ZUTHSM   ZUXSEC    ZUXSA   ZUXSA2   ZUXAFB   ZUTHSM  ZUXSA ZUXSA2 ZUXAFB ZUTPSM  ZUTAU  ZULRSM  ZUALR
    0  0.20718  0.20718
    1  0.11707  0.11707  0.11707  0.11707  0.11707   0.2404 0.2404 0.2404 0.2404
    2  0.11714  0.11714  0.11715  0.11715  0.11714   0.2403 0.2403 0.2403 0.2403
    3  0.11696  0.11696  0.11696  0.11696  0.11696   0.2405 0.2405 0.2405 0.2405-0.1520-0.1520
    4  0.39006  0.39006  0.39006                     0.2003 0.2003                             0.2078 0.0000
    5  0.48082  0.48082  0.48082                     0.1518 0.1518                             0.1962 2.1932
    6  0.39002  0.39002  0.39002                     0.2006 0.2006                             0.2078 2.1932
    7  0.48083  0.48083  0.48083                     0.1518 0.1518                             0.1962 2.0196
    8  0.00000  0.00000  0.00000                     0.0000 0.0000                             0.0000 2.1951
    9  0.47229  0.47229  0.47229                     0.1546 0.1546                             0.1981 0.0000
   10  2.21402  2.21402                                                                        0.2007 0.0000
   11  0.19403  0.19401  0.19400  0.19400            0.6379 0.6381 0.6381


  SQRT(S) =   140.

      <-------------- Cross Section -------------->  <------- Asymmetry ------->  <--Tau_Pol-->  <----A_LR--->
 INDF   ZUTHSM   ZUXSEC    ZUXSA   ZUXSA2   ZUXAFB   ZUTHSM  ZUXSA ZUXSA2 ZUXAFB ZUTPSM  ZUTAU  ZULRSM  ZUALR
    0  0.02252  0.02252
    1  0.01755  0.01755  0.01755  0.01755  0.01755   0.2920 0.2920 0.2920 0.2920
    2  0.01758  0.01758  0.01758  0.01758  0.01758   0.2917 0.2917 0.2917 0.2917
    3  0.01757  0.01757  0.01757  0.01757  0.01757   0.2917 0.2917 0.2917 0.2917-0.1171-0.1171
    4  0.04846  0.04846  0.04846                     0.2478 0.2478                             0.2115 0.0000
    5  0.05371  0.05371  0.05371                     0.1693 0.1694                             0.2045 2.1932
    6  0.04847  0.04847  0.04847                     0.2480 0.2480                             0.2115 2.1932
    7  0.05371  0.05371  0.05371                     0.1693 0.1694                             0.2045 2.0196
    8  0.00000  0.00000  0.00000                     0.0000 0.0000                             0.0000 2.1951
    9  0.05302  0.05302  0.05302                     0.1747 0.1747                             0.2107 0.0000
   10  0.25737  0.25691                                                                        0.2083 0.0000
   11  0.08494  0.08497  0.08498  0.08498            0.8995 0.8996 0.8996


  SQRT(S) =   175.

      <-------------- Cross Section -------------->  <------- Asymmetry ------->  <--Tau_Pol-->  <----A_LR--->
 INDF   ZUTHSM   ZUXSEC    ZUXSA   ZUXSA2   ZUXAFB   ZUTHSM  ZUXSA ZUXSA2 ZUXAFB ZUTPSM  ZUTAU  ZULRSM  ZUALR
    0  0.01078  0.01078
    1  0.00939  0.00939  0.00939  0.00939  0.00939   0.2815 0.2815 0.2815 0.2815
    2  0.00941  0.00941  0.00941  0.00941  0.00941   0.2810 0.2810 0.2810 0.2810
    3  0.00940  0.00940  0.00941  0.00941  0.00940   0.2810 0.2810 0.2810 0.2810-0.1006-0.1006
    4  0.02428  0.02428  0.02428                     0.2444 0.2444                             0.2016 0.0000
    5  0.02587  0.02587  0.02587                     0.1617 0.1618                             0.1953 2.1932
    6  0.02428  0.02428  0.02428                     0.2445 0.2445                             0.2016 2.1932
    7  0.02587  0.02587  0.02587                     0.1617 0.1618                             0.1953 2.0196
    8  0.00000  0.00000  0.00000                     0.0000 0.0000                             0.0000 2.1951
    9  0.02561  0.02561  0.02561                     0.1693 0.1693                             0.2063 0.0000
   10  0.12590  0.12545                                                                        0.1998 0.0000
   11  0.05490  0.05491  0.05491  0.05491            0.9019 0.9021 0.9021

\end{Verbatim}
\end{minipage}
\end{turn}


\eqnzero
\section{Summary}
The description of the \zf\ package has been updated to version 6.42.
Besides a short introduction to the physics contents of the program
additions, we reproduce, in Appendices,  also the technical details
like user flags and interface subroutines as complete as necessary for
a convenient use of the program.

\zf\ covers most of the radiative corrections of a practical
relevance in the foreseeable future. 
Yet, several corrections are still needed  to close the two loop program of
data analysis with \zf:
\begin{itemize} 
\item  cross section (or decay rate) asymmetries, with the $Z$  boson
  coupling to light  quarks  or  $b$ quarks;
\item electroweak corrections to the $Z$  boson width;
\item  bosonic electroweak  corrections  to all  asymmetries and the
  effective weak mixing angle.
\end{itemize}
The inclusion of  the two loop electroweak corrections  to the $Zb\bar{b}$
vertex will additionally require to account for the  mass of the $b$
quark in the one loop electroweak  corrections.  
Nonetheless, it was checked  that the one loop corrections  with a
massive $b$ quark would give negligible effects for the accuracy
reached at LEP. 
 
Some bug reports concerning \zf\ versions between version 6.21 and
version 6.42 will be found at the webpage \cite{zfitter-support-page}.


\ack
For many years, the support of the user community of \zf\ by the
authors of the program was substantial for the numerous applications
of the program by the LEP collaborations, the LEP~EWWG (LEP
Electroweak Working Group), and many other user groups.

We greatly appreciate the readyness of those authors of \zf\ who
decided not to join the \zf\ support group,
Dima Bardin, Pena Christova, Mark Jack, Lida Kalinovskaya, 
Alexandre Olshevski, 
to transfer the support of the program with full responsibility to the
\zf\ support group, thus allowing the program to survive in a
rapidly changing physics world.

We would like to thank Georg Weiglein for discussions. 

A.A. 
is grateful for financial support by RFBR grant 04-02-17192.
The work  of M.C. 
was supported in part by TMR under EC-contract No. HPRN-CT-2002-00311
(EURIDICE).
M.A. and M.C. were supported by 
the Polish State Committee for Scientific Research (KBN)
  for the research project in years 2004-2005, and also 
by the Sofja Kovalevskaja Award of the Alexander von Humboldt
Foundation sponsored by the German Federal Ministry of Education and
Research. 
M.A., M.C. and T.R. 
were supported by European's 5-th Framework under contract
No. HPRN--CT--2000--00149 (Physics at Colliders)
and by Deutsche Forschungsgemeinschaft under contract SFB/TR 9--03.


\appendix
\eqnzero
\section{{\tt DIZET} user guide\label{dizetug}}
\eqnzero
This Appendix describes technical details of the \dz\ package.
Not all of them have been influenced by the updates over the years.
For the convenience of the user, we nevertheless decided to give a
complete overview and will repeat a substantial part of the material,
which was already presented in Section 4.1 of  \cite{Bardin:1999yd}. 
\subsection{Structure of {\tt DIZET}}
A first call of subroutine {\tt DIZET} returns
various pseudo-observables, the $\wb$-boson mass,
weak mixing angles, the $\zb$-boson width, the $\wb$-boson width and
other quantities.
After the first call to {\tt DIZET}, several subroutines of  {\tt DIZET}
might be used for the calculation of form factors and couplings.
This is described in Section 4.1.1 of \cite{Bardin:1999yd}. 
\subsection{Input and output of {\tt DIZET}\label{iodizet}}
The {\tt DIZET} argument list contains
Input, Output and Mixed (I/O) types of arguments:
\\ \\
\begin{small}
\centerline{
{\fbox
{
{\tt 
CALL DIZET(NPAR,AMW,AMZ,AMT,AMH,DAL5H,ALQED,ALSTR,ALSTRT,ZPAR,PARTZ,PARTW)}}}
}
\end{small}
\subsubsection{Input and I/O parameters to be set by the user}
{\bf Input:}
\begin{description}
\item[] {\tt NPAR(1:25), INTEGER*4} vector of flags
\item[] {\tt AMT} = $\mtl$ -- $\ft$-quark mass
\item[] {\tt AMH} = $\mhl$ -- Higgs boson mass 
\item[] {\tt ALSTR} = $\als(\mzs)$ -- strong coupling at $\sman = \mzs$
\end{description}
{\bf I/O:}
\begin{description}
\item[] {\tt AMW} = $\mwl$, $\wb$ boson mass, input if {\tt NPAR(4)} = 2,3,
                  but is being calculated for {\tt NPAR(4)} = 1
\item[] {\tt AMZ} = $\mzl$, $\zb$ boson mass, input if {\tt NPAR(4)} = 
1,3,
                  but is being calculated for {\tt NPAR(4)} = 2 
\item[] {\tt DAL5H} = $\dalhv$, hadronic vacuum polarization
\end{description}
The $\mzs$, $\mws$, and $\dalhv$ cannot be assigned by a parameter 
statement (input/output variables).
\subsubsection{Output of the {\tt DIZET} package}
\begin{description}
\item[{\tt ALQED~}] = $\alpha(\mzs)$, calculated from $\dalhv$, see
description of flag {\tt ALEM} in \subsect{dizetflags}
\item[{\tt ALSTRT}] = $\als(\mts)$         
\item[{\tt ZPAR(1)}] = {\tt DR}=$\dr$, the loop corrections to the
muon decay constant
\item[{\tt ZPAR(2)}] = {\tt DRREM} = $\dr_{\rm{rem}}$, the remainder
contribution $\ord{\alpha}$ 
\item[{\tt ZPAR(3)}] = {\tt SW2} = $\siws$, squared of sine of the weak mixing
angle defined by weak boson masses   
\item[{\tt ZPAR(4)}] = {\tt GMUC} = $\gf$, muon decay constant, if
{\tt NPAR(4)} = 1,2, is set in {\tt CONST1} depending on flag
                                   {\tt NPAR(20)}, see \subsect{dizetflags}. 
(It should be calculated if {\tt NPAR(4)}=3 from $\mzl,\mwl$, but 
then it will deviate from the experimental value.)   
\item[{\tt ZPAR(5-14)}] -- stores effective sines for all partial 
$\zb$-decay channels:
\begin{description}
\item[$~~~~$ ] $~$5 -- neutrino
\item[$~~~~$ ] $~$6 -- electron
\item[$~~~~$ ] $~$7 -- muon
\item[$~~~~$ ] $~$8 -- $\tau$ lepton
\item[$~~~~$ ] $~$9 -- up quark
\item[$~~~~$ ] 10 -- down quark
\item[$~~~~$ ] 11 -- charm quark
\item[$~~~~$ ] 12 -- strange quark
\item[$~~~~$ ] 13 -- top quark (presently equal to up quark)
\item[$~~~~$ ] 14 -- bottom quark
\end{description}
\item[{\tt ZPAR(15)}] = {\tt ALPHST} $\equiv \als(\mzs)$  
\item[{\tt ZPAR(16-30)}] = {\tt QCDCOR(0-14)},
{\tt QCDCOR(I)} -- array of QCD correction factors for quark production 
processes and/or $\zb$ boson partial width (channel $i$) into quarks.
Enumeration as follows:
\renewcommand{\arraystretch}{1.1}
\bqa
\begin{array}{lcl}
\mbox{{\tt QCDCOR(0)} }&=&1\\
\mbox{{\tt QCDCOR(1)} }&=&R^{\fu}_{\sss{V}}(\mzs)\\
\mbox{{\tt QCDCOR(2)} }&=&R^{\fu}_{\sss{A}}(\mzs)\\
\mbox{{\tt QCDCOR(3)} }&=&R^{\fd}_{\sss{V}}(\mzs)\\
\mbox{{\tt QCDCOR(4)} }&=&R^{\fd}_{\sss{A}}(\mzs)\\
\mbox{{\tt QCDCOR(5)} }&=&R^{\fc}_{\sss{V}}(\mzs)\\
\mbox{{\tt QCDCOR(6)} }&=&R^{\fc}_{\sss{A}}(\mzs)\\
\mbox{{\tt QCDCOR(7)} }&=&R^{\fs}_{\sss{V}}(\mzs)\\
\mbox{{\tt QCDCOR(8)} }&=&R^{\fs}_{\sss{A}}(\mzs)\\
\mbox{{\tt QCDCOR(9)} }&=&R^{\fu}_{\sss{V}}(\mzs)\quad\mbox{foreseen for}\quad\ft\bart\mbox{-channel}\\
\mbox{{\tt QCDCOR(10)}}&=&R^{\fu}_{\sss{A}}(\mzs)\quad\mbox{foreseen for}\quad\ft\bart\mbox{-channel}\\
\mbox{{\tt QCDCOR(11)}}&=&R^{\fb}_{\sss{V}}(\mzs)\\
\mbox{{\tt QCDCOR(12)}}&=&R^{\fb}_{\sss{A}}(\mzs)\\
\mbox{{\tt QCDCOR(13)}}&=&R^{\sss{S}}_{\sss{V}} 
\mbox{--singlet vector correction}\\ 
\mbox{{\tt QCDCOR(14)}}&=&f_1,\quad\mbox{corrections to}\quad A_{FB}
\mbox{\cite{Arbuzov:1992pr}}\\
\end{array}
\label{qcdcor_fst}
\eqa
\item[{\tt PARTZ(I)}] -- array of partial decay widths of the $\zb$-boson: \begin{description}
\item[$~~~~$ ]{\tt I} = $~$0   neutrino
\item[$~~~~$ ]{\tt I} = $~$1   electron
\item[$~~~~$ ]{\tt I} = $~$2   muon    
\item[$~~~~$ ]{\tt I} = $~$3   tau     
\item[$~~~~$ ]{\tt I} = $~$4   up      
\item[$~~~~$ ]{\tt I} = $~$5   down    
\item[$~~~~$ ]{\tt I} = $~$6   charm   
\item[$~~~~$ ]{\tt I} = $~$7   strange 
\item[$~~~~$ ]{\tt I} = $~$8   top (foreseen, not realized)
\item[$~~~~$ ]{\tt I} = $~$9   bottom             
\item[$~~~~$ ]{\tt I} = 10  hadrons            
\item[$~~~~$ ]{\tt I} = 11   total  
\end{description}
\item[{\tt PARTW(I)}] -- array of partial decay widths of the 
$\wb$-boson\footnote{The calculation of the $\wb$
width~\cite{Bardin:1986fi} 
follows the same principles as that of the
$\zb$ width and is realized in subroutine {\tt ZWRATE} of {\tt DIZET}.
Since the $\wb$ width is not that important for the description of
fermion pair production, we do not go into details.}
for the channels:
\begin{description}
\item[$~~~~$ ]{\tt I} = 1 one leptonic
\item[$~~~~$ ]{\tt I} = 2 one quarkonic
\item[$~~~~$ ]{\tt I} = 3 totoal
\end{description}
\end{description}

\subsection{The flags used by {\tt DIZET} \label{dizetflags}}
Since the {\tt DIZET} package may be used as  stand-alone in order
to compute POs 
we present here a short description of all flags in  
{\tt DIZET}. 
The flag values 
must be filled in vector {\tt NPAR(1:25)} by the the user.
Most of these flags overlap with the flags set in user subroutine
{\tt ZUFLAG} called by  {\tt ZFITTER},
however, in the stand-alone mode {\tt ZUFLAG} need not be
called. 
We will show the correspondence between the flag names {\tt CHFLAG}
and the flag values {\tt IVALUE} used inside {\tt DIZET},
called with
\\ 
\centerline{$ 
\fbox
{\tt CALL ZUFLAG('CHFLAG', IVALUE)}$\,.}
\\
The description is given in the order of the vector {\tt NPAR(1:25)}.
Flag values marked as {\bf presently not supported} are not recommended.
For instance, they may be chosen for backward compatibility with
respect to earlier versions of the code.

\begin{description}       
\item[{\tt NPAR(1)} = {\tt IHVP}] $\rightarrow\;\;$ 
{\tt ZUFLAG('VPOL',IHVP)} --
Handling of hadronic vacuum polarization:
 \begin{description}
  \item[{\tt IHVP} = 1]  (default) by the parameterization 
                       of~\cite{Eidelman:1995ny}
  \item[{\tt IHVP} = 2]  by effective quark masses
of~\cite{Jegerlehner:1991dq,Jegerlehner:1991ed}~~~{\bf presently not} 
  \item[{\tt IHVP} = 3]  by the parameterization 
                       of~\cite{Burkhardt:1989ky} $\hspace{11mm}$ 
                       {\bf supported}
 \end{description}
\end{description}
\begin{description}
\item[{\tt NPAR(2)} = {\tt IAMT4}] $ \rightarrow $ 
{\tt ZUFLAG('AMT4',IAMT4)} --
Re-summation of the leading $\ord{\gf\mts}$ electroweak corrections,
see \sect{changes}:
 \begin{description}
  \item[{\tt IAMT4} = 0] no re-summation

  \item[{\tt IAMT4} = 1] with re-summation recipe of~\cite{Consoli:1989pc}~~
                      {\bf presently}    
  \item[{\tt IAMT4} = 2] with re-summation recipe of~\cite{Halzen:1991je}  
                      $\hspace{7mm}$ {\bf not}     
  \item[{\tt IAMT4} = 3] with re-summation recipe of~\cite{Fanchiotti:1991kc}~
                      $\hspace{1mm}$ {\bf supported}
  \item[{\tt IAMT4} = 4] (default) with two-loop sub-leading corrections
                       and re-summation recipe 
  of~\cite{Degrassi:1994a0,Degrassi:1995ae,Degrassi:1995mc,%
Degrassi:1996mg,Degrassi:1996ZZ,Degrassi:1999jd}  
  \item[{\tt IAMT4} = 5] with  fermionic two-loop corrections to $\mwl$
                      according to~\cite{Freitas:2002ja,Freitas:2000gg,Freitas:2000nv}
 \item[{\tt IAMT4} = 6] with complete two-loop corrections to $\mwl$
                       \cite{Awramik:2003rn} and
                       fermionic two-loop corrections to
                       $\sin^2 \theta^{\rm lept}_{\rm eff}$ \cite{Awramik:2004ge}
 \end{description}
\end{description}
\begin{description}
\item[{\tt NPAR(3)} = {\tt IQCD}] $ \rightarrow $ 
{\tt ZUFLAG('QCDC',IQCD)}
-- Handling of internal QCD corrections of order $\ord{\alpha\als}$:
\begin{description}
\item[{\tt IQCD} = 0]  no internal QCD corrections 
\item[{\tt IQCD} = 1]  by Taylor expansions (fast option)
of~\cite{Bardin:1989aa} 
\item[{\tt IQCD} = 2]  by exact formulae of~\cite{Bardin:1989aa}
\item[{\tt IQCD} = 3]  (default) by exact formulae of~\cite{Kniehl:1990yc}  
\end{description}
\end{description}
\begin{description}
\item[{\tt NPAR(4)} = {\tt IMOMS}]
-- Choice of two input/output parameters from the three parameters
$\big\{\gf,$ $\mzl,$ $\mwl\bigr\}$:   
 \begin{description}
  \item[{\tt IMOMS} = 1] (default) input $\gf,\mzl$; output $\mwl$),
   see~\eqn{wmass}   
  \item[{\tt IMOMS} = 2] input $\gf,\mwl$; output $\mzl$,  
  \item[{\tt IMOMS} = 3] input $\mzl,\mwl$; output $\gf$, {\bf foreseen, not
  realized} 
 \end{description}
\end{description}
\begin{description}
\item[{\tt NPAR(5)} = {\tt IMASS}]
-- Handling of hadronic vacuum polarization in $\dr$; for tests only:
 \begin{description}
  \item[{\tt IMASS} = 0] (default) uses a fit to data 
  \item[{\tt IMASS} = 1] uses effective quark masses 
 \end{description}
\end{description}
\begin{description}
\item[{\tt NPAR(6)} = {\tt ISCRE}] $\rightarrow$ 
{\tt ZUFLAG('SCRE',ISCRE)}
-- Choice of the scale of the two-loop remainder terms of $\dr$
with the aid of a conversion factor $f$, for details see
\cite{Bardin:1999yd}. 
\begin{description}
\item[{\tt ISCRE} = 0] (default) scale of the remainder terms is $\Ksc=1$ 
\item[{\tt ISCRE} = 1] scale of the remainder terms is $\Ksc=f^2$ 
\item[{\tt ISCRE} = 2] scale of the remainder terms is
$\Ksc={\ds{\frac{1}{f^2}}}$ 
  \end{description}
\end{description}
\begin{description}
\item[{\tt NPAR(7)} = {\tt IALEM}] $ \rightarrow $
{\tt ZUFLAG('ALEM',IALEM)}
-- Controls the usage of $\alpha(\mzs)$, see flowchart in
\cite{Bardin:1999yd}. 
Inside {\tt DIZET}, however, its meaning is limited:
 \begin{description}
  \item[{\tt IALEM} = 0 or 2] $\dalhv$ must be supplied 
   by the user as input to the {\tt DIZET} package        
  \item[{\tt IALEM} = 1 or 3] $\dalhv$ is calculated by the program       
                        using a parameterization {\tt IHVP} (default: IALEM=3) 
 \end{description}
\end{description}
For details see the complete discussion about this flag in 
Sections 2.8 and 4.2.2 of \cite{Bardin:1999yd}.
\begin{description}
\item[{\tt NPAR(8)} = {\tt IMASK}]
-- Historical relict of earlier versions. {\bf Presently unused.}
\end{description}
\begin{description}
\item[{\tt NPAR(9)} = {\tt ISCAL}]  $ \rightarrow $ 
{\tt ZUFLAG('SCAL',ISCAL)}
-- Choice of the scale of $\als(\xi\mtl)$:
 \begin{description}
  \item[{\tt ISCAL} = 0] (default) exact {\tt AFMT} 
                       correction \cite{Avdeev:1994db}
  \item[{\tt ISCAL} = 1]\hspace{-2mm}{\bf ,2,3} options used
in~\cite{Kniehl:1995yr},  
        {\bf presently not supported}
  \item[{\tt ISCAL} = 4] Sirlin's scale $\xi= 0.248$ ~\cite{Sirlin:1995yr}
 \end{description}
\end{description}
\begin{description}
\item[{\tt NPAR(10)} =  {\tt IBARB}]  $ \rightarrow $
{\tt ZUFLAG('BARB',IBARB)}
-- Handling of leading $\ord{\gfs\mtq}$ corrections: 
 \begin{description}
  \item[{\tt IBARB} = 0] corrections are not included 
  \item[{\tt IBARB} = 1]  corrections are applied in the 
                    limiting case: Higgs mass negligible with
                    respect to the top mass, \cite{vanderBij:1987hy}
  \item[{\tt IBARB} = 2] (default) analytic results of 
\cite{Barbieri:1993ra} approximated by a polynomial 
\cite{Barbieri:1999bbo}
 \end{description}
These options are inactive for {\tt AMT4} = 4.
\end{description}
\begin{description}
\item[{\tt NPAR(11)} = {\tt IFTJR}] $ \rightarrow $
{\tt ZUFLAG('FTJR',IFTJR)} 
-- Treatment of $\ord{\gf\als\mts}$ {\tt FTJR} corrections
\cite{Fleischer:1992fq}, 
 \begin{description}  
  \item[{\tt IFTJR} = 0]  without {\tt FTJR} corrections
  \item[{\tt IFTJR} = 1]\hspace{-2mm}{\bf ,2}  with~~{\tt FTJR} corrections (default IFTJR=1)
 \end{description}
\end{description} 
Inside {\tt DIZET} its meaning is limited.
See complete discussion about this flag in \cite{Bardin:1999yd}.
\begin{description}
\item[{\tt NPAR(12)} = {\tt IFACR}] $ \rightarrow $
{\tt ZUFLAG('EXPR',IFACR)}
-- Realizes different expansions of $\dr$ ~\cite{Degrassi:1996mg,Degrassi:1997ps}: 
\begin{description}
\item[{\tt  IFACR} =  0]  (default) realizes the so-called OMS-I renormalization scheme 
\item[{\tt  IFACR} = 1]  intermediate step from OMS-I to OMS-II renormalization scheme
\item[{\tt  IFACR} = 2]  approaches the spirit of the OMS-II renormalization scheme, a fully expanded option
\end{description}
\end{description}
\begin{description}
\item[{\tt NPAR(13)} = {\tt IFACT}] $ \rightarrow $
{\tt ZUFLAG('EXPF',IFACT)}
-- To simulate theoretical uncertainties different expansions of the formfactors $\rho$ and $\kappa$ are realized in complete analogy to the flag {\tt  IFACR}:
\begin{description}
\item[{\tt  IFACT} = 0] (default)  OMS-I renormalization scheme 
\item[{\tt  IFACT} = 1] intermediate step from OMS-I to OMS-II renormalization scheme
\item[{\tt  IFACT} = 2]  approaches the OMS-II renormalization scheme
\end{description}
\end{description}
\begin{description}
\item[{\tt NPAR(14)} = {\tt IHIGS}] $ \rightarrow $
{\tt ZUFLAG('HIGS',IHIGS)} --
Switch on/off resummation of the leading Higgs contribution:
 \begin{description}
  \item[{\tt IHIGS} = 0] (default) leading Higgs contribution is not re-summed
  \item[{\tt IHIGS} = 1] leading Higgs contribution is re-summed 
 \end{description}
\end{description}
\begin{description}
\item[{\tt NPAR(15)} = {\tt IAFMT}]  $ \rightarrow $
{\tt ZUFLAG('AFMT',IAFMT)}
-- Includes the three-loop corrections {\tt AFMT}$\sim \delta^{\alpha \alpha_s}$ ~\cite{Avdeev:1994db} (see also
description of flag {\tt SCAL}): 
 \begin{description}  
  \item[{\tt IAFMT} = 0 ]  without {\tt AFMT} correction
  \item[{\tt IAFMT} = 1 ]  correction ${\cal{O}}(G_fm_t^2\alpha_s^2)$ is included
  \item[{\tt IAFMT} = 2 ]  corrections ${\cal{O}}(G_fm_t^2\alpha_s^2)$ and  ${\cal{O}}(G_fM_{\zb}^2 \alpha_s^2+\log(m_t^2))$ are included
  \item[{\tt IAFMT} = 3 ]  (default) corrections ${\cal{O}}(G_fm_t^2\alpha_s^2)$, ${\cal{O}}(G_fM_{\zb}^2 \alpha_s^2+\log(m_t^2))$ and ${\cal{O}}(G_fM_{\zb}^2/m^2_t \alpha_s^2)$ are included
 \end{description}
\end{description} 
\begin{description}
\item[{\tt NPAR(16)} = {\tt IEWLC}] -- Treatment of the remainder
terms of $\rho$ and $\kappa$
(used in {\tt ROKAPP} together with obsolete option {\tt AMT4} = 1-3):
 \begin{description}  
  \item[{\tt IEWLC} = 0] all remainders are set equal to zero
  \item[{\tt IEWLC} = 1] (default) standard treatment
 \end{description}
\end{description} 
\begin{description}
\item[{\tt NPAR(17)} = {\tt ICZAK}] $ \rightarrow $
{\tt ZUFLAG('CZAK',ICZAK)} --
Treatment of the {\tt CKHSS} non-factorizable $\ord{\alpha\als}$ corrections
$\Delta_{EW/QCD}$ to the quarkonic width,
$\Gamma_q$~\cite{Czarnecki:1996ei,Harlander:1998zb}:
  \begin{description} 
   \item[{\tt ICZAK} = 0] without {\tt CKHSS} corrections
   \item[{\tt ICZAK} = 1] (default) with {\tt CKHSS} corrections
   \item[{\tt ICZAK} = 2] with {\tt CKHSS} for pseudoobservables and without {\tt CKHSS} in case of realistic observables
  \end{description}
\end{description} 
Inside the weak library {\tt DIZET} the meaning of {\tt ICZAK} is limited. See also discussion about this flag in \subsect{zuflag}.
\begin{description}
\item[{\tt NPAR(18)} = {\tt IHIG2}] $ \rightarrow $
{\tt ZUFLAG('HIG2',IHIG2)} -- Handling of the quadratically enhanced 
two-loop Higgs contributions to $\Delta r$~\cite{vanderBij:1984bw,vanderBij:1984aj}:
 \begin{description}
  \item[{\tt IHIG2} = 0] without Higgs corrections 
  \item[{\tt IHIG2} = 1] with~~~~~Higgs corrections
 \end{description}
\end{description}
\begin{description}
\item[{\tt NPAR(19)} = {\tt IALE2}] $ \rightarrow $
{\tt ZUFLAG('ALE2',IALE2)} --
Treatment of leptonic corrections to $\Delta\alpha$:
 \begin{description}
  \item[{\tt IALE2} = 0]  for backward compatibility with versions up to 
                        v.5.12
  \item[{\tt IALE2} = 1]  with  one-loop corrections
  \item[{\tt IALE2} = 2]  with two-loop corrections~\cite{Kallen:1955ks}
  \item[{\tt IALE2} = 3]  (default) with three-loop corrections~\cite{Steinhauser:1998rq}
 \end{description}
\end{description}
\begin{description}
\item[{\tt NPAR(20)} = {\tt IGFER}]  $ \rightarrow $
{\tt ZUFLAG('GFER',IGFER)} --
Handling of QED corrections to the Fermi constant:
 \begin{description}
  \item[{\tt IGFER} = 0] for backward compatibility with versions
                       up to v.5.12
  \item[{\tt IGFER} = 1] one-loop QED corrections for Fermi constant
                    \cite{Berman:1958,Kinoshita:1958ru,Kallen:1968}
  \item[{\tt IGFER} = 2] two-loop QED corrections for Fermi constant
                    \cite{vanRitbergen:1998yd,vanRitbergen:1998hn}
 \end{description}
\end{description}
\begin{description}
\item[{\tt NPAR(21)} = {\tt IDDZZ}] --  Used in {\tt ZWRATE} for
internal tests: 
 \begin{description}
  \item[{\tt IDDZZ} = 0] {\tt RQCDV(A)} are set to $0$
  \item[{\tt IDDZZ} = 1] (default) standard treatment of FSR QCD corrections
 \end{description}
\end{description}
\begin{description}
\item[{\tt NPAR(22)} = {\tt IAMW2}]  $ \rightarrow $
{\tt ZUFLAG('AMW2',IAMW2)} --
incorporates the fermionic two-loop contributions to the prediction for the W boson mass ~\cite{Freitas:2000gg,Freitas:2000nv,Freitas:2002ja}:
 \begin{description}
  \item[{\tt IAMW2} = 0] (default) no two-loop corrections to $M_{\wb}$ 
  \item[{\tt IAMW2} = 1] with  two-loop corrections to $M_{\wb}$ 
 \end{description}
\end{description}
\begin{description}
\item[{\tt NPAR(23)} = {\tt ISFSR}]  $ \rightarrow $
{\tt ZUFLAG('SFSR',ISFSR)} --
allows to switch the final state radiation 
 \begin{description}
  \item[{\tt ISFSR} =-1] both, QED and QED$\otimes$QCD final state radiation are excluded
  \item[{\tt ISFSR} = 0] final state QED radiation is excluded, QED$\otimes$QCDis included 
  \item[{\tt ISFSR} = 1] (default) both, QED and QED$\otimes$QCD final state radiation are included
 \end{description}
\end{description}
\begin{description}
\item[{\tt NPAR(24)} = {\tt IDMWW}] $ \rightarrow $
{\tt ZUFLAG('DMWW',IDMWW)} -- Simulation of the theoretical error on $\mwl$ for
${\tt AMT4} = 5,6$:
 \begin{description}
 \item[{\tt IDMWW} = -1] minimal value for $\mwl$
 \item[{\tt IDMWW} = 0] (default) no shift on $\mwl$ applied
 \item[{\tt IDMWW} = 1] maximal value for $\mwl$
 \end{description}
\end{description}
\begin{description}
\item[{\tt NPAR(25)} = {\tt IDSWW}] $ \rightarrow $
{\tt ZUFLAG('DSWW',IDSWW)} -- Simulation of the theoretical error on $\sin^2
\theta^{\rm lept}_{\rm eff}$ for ${\tt AMT4} = 6$:
 \begin{description}
 \item[{\tt IDSWW} = -1] minimal value for $\sin^2 \theta^{\rm lept}_{\rm eff}$
 \item[{\tt IDSWW} = 0] (default) no shift on
               $\sin^2 \theta^{\rm lept}_{\rm eff}$ applied
 \item[{\tt IDSWW} = 1] maximal value for $\sin^2 \theta^{\rm lept}_{\rm eff}$
 \end{description}
\end{description}
%
\subsection{\label{xfotf3}Calculation of $\alpha(\sman)$. Function {\tt
XFOTF3}} 
the running QED coupling at scale $\sman$ is calculated with function
{\tt XFOTF3} as follows:
\bq
\alpha(\sman)=\frac{\alpha}
{\ds{1-
\frac{\alpha}{4\pi}{\tt DREAL(XFOTF3(IALEM,IALE2,IHVP,IQCD,1,DAL5H,-S))}}}\;.
\eq

 
\eqnzero
\section{\zf\ user guide\label{zfguide}}
\eqnzero
This Appendix describes technical details of the \zf\ package.
Not all of them have been influenced by the updates over the years.
For the convenience of the user, we nevertheless decided to give a
complete overview and will repeat a substantial part of the material,
which was already presented in Section 4.2 of  \cite{Bardin:1999yd}.

\zf\ is coded in {\tt FORTRAN 77}. 
Double-precision variables have been used throughout the program.
The package consists of the following {\tt FORTRAN} files:
\[
\begin{array}{l}
{\tt zf630\_aux.f }\\
{\tt zfbib6\_40.f }\\
{\tt zfmai6\_42.f }\\
{\tt zfusr6\_42.f }\\
{\tt acol6\_1p.f  }\\
{\tt bcqcdl5\_14.f}\\
{\tt bhang4\_640.f}\\
{\tt bkqcdl5\_14.f}\\
{\tt dizet6\_42.f }\\
{\tt expifi6\_30.f }\\
{\tt funang6\_30.f }\\
{\tt m2tcor5\_11.f}\\
{\tt pairho6\_40.f }\\
{\tt APV\_lib.f}    \\
\end{array}
\]
The following routines are normally called in the initialization phase
of programs using the \zf\ package in the order listed below:
 {\tt ZUINIT}, {\tt ZUFLAG},  {\tt ZUWEAK},  {\tt ZUCUTS},  {\tt ZUINFO}.
An example of different use is described in Section 2.5 of
\cite{Bardin:1999yd}.  
 
\subsection{Subroutine {\tt ZUINIT}\label{zuinit}}
Subroutine {\tt ZUINIT} is used to initialize variables with
their default values.
This routine {\em must} be called before any other \zf\ routine.
 
\SUBR{CALL ZUINIT}
\subsection{Subroutine {\tt ZUFLAG}\label{zuflag}}
Subroutine {\tt ZUFLAG} is used to modify the default
values of flags which control various \zf\ options.
 
\SUBR{CALL ZUFLAG(CHFLAG,IVALUE)}

\noindent \underline{Input Arguments:}
\begin{description}
  \item[\tt CHFLAG] is the character identifier of a \zf\ flag.
                
  \item[\tt IVALUE] is the value of the flag. See \tbn{tab:xxxx} for
                        a list of the defaults. 
\end{description}
Possible combinations of {\tt CHFLAG} and {\tt IVALUE} are listed below:\footnote{It is worth noting that
not for all flags 
the default value is necessarily the preferred value.
A typical example is flag {\tt FINR}, distinguishing two different treatments
of FSR, which are relevant in different experimental setups.}

In \tbn{tab:xxxx} an overview over all flags used in
{\tt DIZET} and {\tt ZFITTER} is given. 
The {\tt DIZET} flags  (vector 
{\tt NPAR(1:25)} in {\tt DIZET} corresponds to   {\tt NPARD(1:25)} in  {\tt ZFITTER})
are described in \subsect{dizetug}.

\begin{description} 
  \item[\tt AFBC] --
  Controls the calculation of the forward
  backward asymmetry for interfaces {\tt ZUTHSM}, {\tt ZUXSA},
  {\tt ZUXSA2}, and {\tt ZUXAFB}:
  \begin{description}
    \item[{\tt IVALUE} = 0]
    asymmetry calculation is inhibited (can speed up the program
    if asymmetries are not desired)
    \item[{\tt IVALUE} = 1]
    (default) both cross-section and asymmetry calculations are done
  \end{description}
\end{description}
 
\begin{description} 
  \item[\tt AFMT] -- see {\tt NPAR(15)} in subsection \ref{dizetflags}
\end{description}

\begin{description} 
  \item[\tt ALEM] --
  Controls the treatment of the running QED coupling $\alpha(\sman)$:
  \begin{description}
  \item[{\tt IVALUE} = 0 or 2] $\dalhv$ must be supplied 
   by the user as input to the {\tt DIZET} package;
   using this input {\tt DIZET} calculates {\tt ALQED} = $\alpha(\mzs)$
  \item[{\tt IVALUE} = 1 or 3] $\dalhv$ and $\alpha(\mzs)$ are calculated 
   by the program using a parameterization {\tt IHVP}.
  \end{description}
\end{description}

   The scale of $\alpha($scale$)$ is governed in addition by the flag 
   {\tt CONV}, see description below and the flowchart, Figure 6 of
   \cite{Bardin:1999yd}. 
   Values {\tt ALEM} = 0,1 are accessible only at {\tt CONV} = 0.  
   Then for {\tt ALEM} = 0,1 $\alpha(\mzs)$ 
   and for {\tt ALEM} = 2,3  $\alpha(\sman)$ are calculated.             
   Values {\tt ALEM} = 2,3 are accessible for {\tt CONV} = 0,1,2.  
   Then for {\tt CONV} = 0 $\alpha(\sman)$ 
   and for {\tt CONV} = 1,2 $\alpha(\smanp)$ are calculated. 
   Recommended values: {\tt ALEM} = 2,3.                    

\begin{description} 
  \item[\tt ALE2] -- see {\tt NPAR(19)} in subsection \ref{dizetflags}
\end{description}

\begin{description} 
  \item[\tt AMT4] -- see {\tt NPAR(2)} in subsection \ref{dizetflags}
\end{description}

\begin{description} 
  \item[\tt ASCR] -- is a hidden flag that handles the treatment of contributions to $A_{FB}$:  
 \begin{description} 
  \item[\tt IVALUE = 0] treatment as in versions 5 up to version 6.23
  \item[\tt IVALUE = 1] (default) new (and very old) treatment
 \end{description}
\end{description}

\begin{description} 
  \item[\tt BARB] -- see {\tt NPAR(10)} in subsection \ref{dizetflags}
\end{description}

\begin{description} 
  \item[\tt BORN] --
  Controls calculation of QED and Born observables:
  \begin{description}
    \item[{\tt IVALUE} = 0]
    (default) QED convoluted observables
    \item[{\tt IVALUE} = 1]
    electroweak observables corrected by Improved Born Approximation
  \end{description}
\end{description}
 
\begin{description}
  \item[\tt BOXD] --
  Determines calculation of $ZZ$ and $WW$ box
  contributions,
  (see Section 3.3.2 of \cite{Bardin:1999yd}):
  \begin{description}
    \item[{\tt IVALUE} = 0]
    no box contributions are calculated
    \item[{\tt IVALUE} = 1]
    (default) the boxes are calculated as additive separate contribution
    to the cross-section  
    \item[{\tt IVALUE} = 2]
    box contributions are added to all four form factors    
  \end{description}
\end{description}

\begin{description} 
  \item[\tt CONV] --
  Controls the energy scale of running $\alpha$ and EWRC,
  see  Figure 6 of \cite{Bardin:1999yd}:
  \begin{description} 
    \item[{\tt IVALUE} = 0]
    $\alpha(s)$
    \item[{\tt IVALUE} = 1]
    (default) $\alpha(\smanp)$ convoluted
    \item[{\tt IVALUE} = 2]
    both electroweak radiative correction and $\als$ are convoluted
  \end{description}
\end{description}

\begin{description} 
  \item[\tt CZAK] --
  Treatment of {\tt CKHSS} non-factorized corrections,
  \cite{Czarnecki:1996ei},~\cite{Harlander:1998zb},~see~ Figure 6 of
  \cite{Bardin:1999yd}: 
  \begin{description} 
    \item[{\tt IVALUE} = 0] without {\tt CKHSS} corrections 
    \item[{\tt IVALUE} = 1] (default) with {\tt CKHSS} corrections everywhere
    \item[{\tt IVALUE} = 2] {\tt CKHSS} corrections are taken into account only       in POs, this option is used for tests only
  \end{description}
\end{description}

\begin{description} 
  \item[\tt DIAG] --
  Selects type of diagrams taken into account:
  \begin{description} 
    \item[{\tt IVALUE} = --1]
    only $\zb$- exchange diagrams are taken into account 
    \item[{\tt IVALUE} = 0]
    $\zb$ and $\ph$ - exchange diagrams are taken into account 
    \item[{\tt IVALUE} = 1]
    (default) $\zb$ and $\ph$ exchange and $\zb\ph$ interference are 
    included
  \end{description}
\end{description}

\begin{description} 
  \item[\tt EXPF] -- see {\tt NPAR(13)} in subsection \ref{dizetflags}
\end{description}

\begin{description} 
  \item[\tt EXPR] -- see {\tt NPAR(12)} in subsection \ref{dizetflags}
\end{description}

\begin{description} 
  \item[\tt ENUE] -- Treatment of the improved Born approximation for
    the process $\fep\fem \rightarrow \fnue \bar{\nu_e}$. This option
    is only available via subroutine {\tt COSCUT} in {\tt
      zfbib6\_34.f} and using {\tt zfEENN\_34.f}.  
This flag is {\em presently not supported}, see also Section \ref{sec-nunu}.  
\begin{description} 
 \item[{\tt IVALUE} = --1] s-channel only    
 \item[{\tt IVALUE} = 0]   s- and t-channels          
 \item[{\tt IVALUE} = 1]  (default) s+t complete with s-t interference 
  \end{description}
\end{description}

\begin{description} 
  \item[\tt DMWW] -- see {\tt NPAR(24)} in subsection \ref{dizetflags}  
\end{description}

\begin{description} 
  \item[\tt DSWW] -- see {\tt NPAR(25)} in subsection \ref{dizetflags}  
\end{description}

\begin{description} 
  \item[\tt FBHO] -- treatment of second order corrections to angular distributions and  $A_{FB}$
  \begin{description}
    \item[{\tt IVALUE} = 0] (default) treatment as in version 6.21 and before
    \item[{\tt IVALUE} = 1] modified treatment; photonic radiative corrections are improved and ${\cal{O}}(\alpha^2)$ contributions from pairs are included in leading-log approximation, see Section \ref{sec-qed-1}
   \end{description}
\end{description}

\begin{description} 
  \item[\tt FINR] --
  Controls the calculation of final-state radiation,
  \begin{description}
    \item[{\tt IVALUE} = --1]
    final-state QED and QCD correction are not applied;
    \item[{\tt IVALUE} = 0]
    by $\smanp$ cut,
    final-state QED correction is described with the factor
    $1 + 3 \alpha(\sman) / (4 \pi ) \qfs$
    \item[{\tt IVALUE} = 1]
    (default) $M^{2}_{ff}$ cut,
    includes complete treatment of final-state radiation
    with common soft-photon exponentiation
  \end{description}
\end{description}
 
\begin{description}
  \item[\tt FOT2] --
  Controls second-order leading log and next-to-leading 
  log QED corrections:
  \begin{description}
    \item[{\tt IVALUE} = --1]
    no initial state radiation QED convolution at all
    \item[{\tt IVALUE} = 0]
    complete $\alpha$ additive radiator
    \item[{\tt IVALUE} = 1]
    with logarithmic hard corrections
    \item[{\tt IVALUE} = 2]
    complete $\alpha^2$ additive radiator 
    \item[{\tt IVALUE} = 3] 
    (default) complete $\alpha^3$ additive radiator 
    \item[{\tt IVALUE} = 4] 
    optional $\alpha^3$ additive radiator for estimation of 
    theoretical errors 
\cite{Bardin:1989qr} 
    \item[{\tt IVALUE} = 5]  
    ``pragmatic'' LLA third order corrections in a factorized
form~\cite{Skrzypek:1992vk} 
  \end{description}
\end{description}

\begin{description}
  \item[\tt FSPP] -- correction due to final state radiation into
    pairs, see Section \ref{sec-qed-4}: 
  \begin{description}
   \item[{\tt IVALUE} = 0] (default) no final state state pair corrections
   \item[{\tt IVALUE} = 1] final state state pair contributions are implemented as additive corrections 
   \item[{\tt IVALUE} = 2]final state state pair contributions are implemented as multiplicative corrections 
 \end{description}
\end{description}

\begin{description}
  \item[\tt FSRS] --
  Final state radiation scale:
  \begin{description}
    \item[{\tt IVALUE} = 0] $\alpha(0)$, preferred for tight cuts
    \item[{\tt IVALUE} = 1] (default) $\alpha(s)$, preferred for loose cuts
  \end{description}
\end{description}

\begin{description} 
  \item[\tt FTJR] --
  Treatment of {\tt FTJR} corrections \cite{Fleischer:1992fq}:
  \begin{description} 
    \item[{\tt IVALUE} = 0] without {\tt FTJR}  corrections 
    \item[{\tt IVALUE} = 1] (default) with {\tt FTJR}  corrections everywhere
    \item[{\tt IVALUE} = 2] {\tt FTJR}  corrections are taken into account only 
                            in POs, the option is used for tests only
  \end{description}
\end{description}

\begin{description} 
  \item[\tt FUNA] --
implementation of higher order photonic corrections to the angular
distribution, see Section \ref{sec-qed-2}: 
  \begin{description} 
    \item[{\tt IVALUE} = 0] (default) no higher order corrections
    \item[{\tt IVALUE} = 1] higher order photonic LLA corrections are included
  \end{description}
\end{description}

\begin{description}
  \item[\tt GAMS] --
  Controls the $\sman$ dependence of ${\cal G}_Z$,
  the \Z-width function, see Section 3.2  of
   \cite{Bardin:1999yd}:
  \begin{description}
    \item[{\tt IVALUE} = 0]
    forces ${\cal G}_Z$ to be constant
    \item[{\tt IVALUE} = 1]
    (default) allows ${\cal G}_Z$
    to vary as a function of $\sman$ \cite{Bardin:1988xt}.
  \end{description}
\end{description}

\begin{description} 
  \item[\tt GFER] -- see {\tt NPAR(20)} in subsection \ref{dizetflags}
\end{description}

\begin{description} 
  \item[\tt HIGS] -- see {\tt NPAR(14)} in subsection \ref{dizetflags}
\end{description}

\begin{description} 
  \item[\tt HIG2] -- see {\tt NPAR(18)} in subsection \ref{dizetflags}
\end{description}

\begin{description}
  \item[\tt INTF] --
  Determines if the ${\cal O} (\alpha)$ initial-final state
  QED interference (IFI) is calculated;  see Section \ref{sec-qed-3}:
  \begin{description}
    \item[{\tt IVALUE} = 0]
    the interference term is ignored
    \item[{\tt IVALUE} = 1]
    (default) with IFI in the $\ord{\alpha}$ 
   \item[{\tt IVALUE} = 2]
    with one-loop IFI corrections and corresponding higher order effects from the exponentiation
  \end{description}
\end{description}

\begin{description}
  \item[\tt IPFC] --
  Pair flavour content for the pair production corrections:
  \begin{description}
    \item[{\tt IVALUE} = 1]
       only electron pairs
    \item[{\tt IVALUE} = 2]       
            only muon pairs
    \item[{\tt IVALUE} = 3]
              only tau-lepton pairs
    \item[{\tt IVALUE} = 4]
              only hadron pairs
    \item[{\tt IVALUE} = 5]
              (default) all channels summed            
    \item[{\tt IVALUE} = 6]
              leptonic pairs (without hadrons)
  \end{description}
\end{description}

\begin{description}
  \item[\tt IPSC] --
 Pair production singlet-channel contributions (works with ISPP = 2): 
  \begin{description}
    \item[{\tt IVALUE} = 0]
         (default) only non-singlet pairs                              
    \item[{\tt IVALUE} = 1]
         LLA singlet pairs according to~\cite{Berends:1988ab}       
    \item[{\tt IVALUE} = 2]
        complete $O(\alpha^2)$ singlet pairs, {\it ibid}
    \item[{\tt IVALUE} = 3]
        singlet pairs up to order $(\alpha L)^3$, {\it ibid}
  \end{description}
\end{description}

\begin{description}
  \item[\tt IPTO] --
    Third (and higher) order pair production contributions
\cite{Arbuzov:2001rt} 
    (works with {\tt ISPP} = 2):              
  \begin{description}
     \item[{\tt IVALUE} = --1] allows to calculate the pure virtual pair contributions separately  
     \item[{\tt IVALUE} = 0]  
 only $O(\alpha^2)$ contributions        
     \item[{\tt IVALUE} = 1] 
  $O(\alpha^3)$ pairs                    
    \item[{\tt IVALUE} = 2]  
  some "non-standard" $O(\alpha^3)$ LLA pairs added       
    \item[{\tt IVALUE} = 3]
(default) $O(\alpha^4)$ LLA electron pairs added                 
  \end{description}
\end{description}

\begin{description}
  \item[\tt ISPP] --
  Treatment of ISR pairs:
  \begin{description}
    \item[{\tt IVALUE} = --1]
    pairs are treated with a ``fudge'' factor as in versions up to v.5.14
    \item[{\tt IVALUE} = 0]
    without ISR pairs
    \item[{\tt IVALUE} = 1]
    with ISR pairs,~\cite{Kniehl:1988id} with a re-weighting 
   \item[{\tt IVALUE} = 2]
   (default) with ISR pairs according to~\cite{Arbuzov:2001rt}
   \item[{\tt IVALUE} = 3]
   with ISR pairs according to~\cite{Jadach:1992aa}
   \item[{\tt IVALUE} = 4]
   with ISR pairs~\cite{Jadach:1992aa} with extended treatment of 
   hadron pair production
  \end{description}
\end{description}
 
\begin{description}
  \item[\tt MISC] --
  Controls the treatment of scaling of $\rho$ in the Model Independent 
  approach, see discussion in \subsect{POCOMM}:
  \begin{description}
    \item[{\tt IVALUE} = 0]
    (default) non-scaled $\rho$'s are used, {\tt AROTFZ}-array
    \item[{\tt IVALUE} = 1]
    scaled $\rho$'s, absorbing imaginary parts, are used, {\tt ARROFZ}-array 
  \end{description}
\end{description}

\begin{description}
  \item[\tt MISD] --
  Controls the $\sman$ dependence of the Model Independent approach
  \begin{description}
    \item[{\tt IVALUE} = 0]
    fixed $\sman = \mzs$ in EWRC, old treatment  
    \item[{\tt IVALUE} = 1]
    (default) ensures equal numbers from all interfaces and for all partial
    channels but {\tt INDF=10} for a large range of $\sqrt{\sman}$
    and for {\tt INDF=10}  
    up to 100 GeV
  \end{description}
\end{description}

\begin{description}
  \item[\tt PART] --
  Controls the calculation of various parts of Bhabha scattering:
  \begin{description}
    \item[{\tt IVALUE} = 0]
    (default) calculation of full Bhabha cross-section and asymmetry
    \item[{\tt IVALUE} = 1] 
    only $\sman$ channel
    \item[{\tt IVALUE} = 2]
    only $\tman$ channel
    \item[{\tt IVALUE} = 3]
    only $\sman-\tman$ interference
  \end{description}
\end{description}

\begin{description}
  \item[\tt POWR] --
  Controls inclusion of final-state fermion masses
  in kinematical factors, see Figure 6 of \cite{Bardin:1999yd}. 
  It acts differently for quarks and leptons. For leptons: 
  \begin{description}
    \item[{\tt IVALUE} = 0]
    final state lepton masses are set equal to zero
    \item[{\tt IVALUE} = 1]
    (default) final state lepton masses are retained
    in all kinematical factors
  \end{description}
  For quarks it is active only for {\tt FINR} = --1 and then: 
  \begin{description}
    \item[{\tt IVALUE} = 0]
    final state quark masses are set equal to zero
    \item[{\tt IVALUE} = 1]
    (default) final state quark masses are set to their running values
    (that is, for $\fc\barc$ and $\fb\barb$ channels)
    and retained in all kinematical factors
  \end{description}
\end{description}

\begin{description}
  \item[\tt PREC] --
  is an integer number which any precision governing any 
  numerical integration is divided by, increasing thereby the numerical   
  precision of computation:
  \begin{description}
    \item[{\tt IVALUE} = 10]
    (default)
    \item[{\tt IVALUE} = 1 -- 99]
    in some cases when some numerical instability 
    while running v.5.10 was registered, it was 
    sufficient to use {\tt PREC} = 3,     
    in some other cases (e.g. with $P_\tau$) only {\tt PREC} = 30 
    solved the instability         
  \end{description}
\end{description}

\begin{description}
  \item[\tt PRNT] --
  Controls {\tt ZUWEAK} printing:
  \begin{description}
    \item[{\tt IVALUE} = 0]
    (default) printing by subroutine {\tt ZUWEAK} is suppressed
    \item[{\tt IVALUE} = 1]
    each call to {\tt ZUWEAK} produces some output
  \end{description}
\end{description}
 
\begin{description} 
  \item[\tt QCDC] -- see {\tt NPAR(3)} in subsection \ref{dizetflags}
\end{description}

\begin{description} 
  \item[\tt SCAL] -- see {\tt NPAR(9)} in subsection \ref{dizetflags}
\end{description}

\begin{description} 
  \item[\tt SCRE] -- see {\tt NPAR(6)} in subsection \ref{dizetflags}
\end{description}

\begin{description} 
  \item[\tt SFSR] -- see {\tt NPAR(23)} in subsection \ref{dizetflags}
\end{description}

\begin{description} 
  \item[\tt TUPV] -- simulates theoretical uncertainties in APV, see Sect. \ref{sect:apv}:
  \begin{description} 
    \item[\tt IVALUE = 1] (default)
    \item[\tt IVALUE = 2,3] for variation, see Sect. \ref{sect:apv}
  \end{description}
\end{description}

\begin{description} 
  \item[\tt VPOL] -- see {\tt NPAR(1)} in subsection \ref{dizetflags}
\end{description}

\begin{description}
  \item[\tt WEAK] --
  Determines if the weak loop calculations are to be performed
  \begin{description}
    \item[{\tt IVALUE} = --1] is only valid for {\tt INDF}=$-$1 and represents the the options {\tt WEAK=1,2} without electroweak corrections for \wb\ boson exchange in the t-channel
    \item[{\tt IVALUE} = 0]
    no weak loop corrections to the cross-sections are calculated
    and weak parameters are forced to their Born values,  i.e.
    $\rho_{ef} = \kappa_{e,f,ef} = 1$
    \item[{\tt IVALUE} = 1]
    (default) weak loop corrections to the cross-sections are
    calculated
    \item[{\tt IVALUE} = 2] weak loop corrections are calculated but some higher order corrections that do not propagate via {\tt DIZET} are swiched off ({\tt ADDIME, ADDIMF}; these corrections are small at LEP2 and they are not included by using {\tt DIZET} with other codes
  \end{description}
\end{description}
\subsection{Subroutine {\tt ZUWEAK}\label{zuweak}}
Subroutine {\tt ZUWEAK} is used to perform the weak sector calculations.
These are done internally with {\tt DIZET}, see~\sect{dizetug}.
The routine calculates a number of important electroweak parameters
which are stored in common blocks for later use (see Section \ref{POCOMM}).
If any \zf\ flag has to be modified this must be done before
calling {\tt ZUWEAK}.
 
\SUBR{CALL ZUWEAK(ZMASS,TMASS,HMASS,DAL5H,ALFAS)}

\noindent \underline{Input Arguments:}
\smallskip
\begin{description}
  \item[\tt ZMASS] is the \Z\  mass  $\mzl$  in GeV.
  \item[\tt TMASS] is the top quark mass  $\mtl$  in GeV, [10-400].
  \item[\tt HMASS] is the Higgs mass  $\mhl$  in GeV, [10-1000].
  \item[\tt DAL5H] is the value of $\dalhv$.
  \item[\tt ALFAS] is the value of the strong coupling constant
     $\als$  at $q^2 = \mzs $ 
(see factors {\tt QCDCOR} in \tbn{qcdcor_fst}).
\end{description}
 
Computing time may be saved by performing
weak sector calculations only once during the initialization of the \zf\
package.
This is possible since weak parameters are nearly
independent of $s$ near the \Z\ peak, \EG\ $\sim \ln s/\mzs$.
However, the incredible precision of \LEPI\ data forced us to give
up this option, see description of flag {\tt MISD}.
\subsection{Subroutine {\tt ZUCUTS}\label{zucuts}}
 
Subroutine {\tt ZUCUTS} is used to define kinematic and geometric cuts
for each fermion channel:
it selects
the appropriate QED calculational {\em chain}.
 
\SUBR{CALL  ZUCUTS(INDF,ICUT,ACOL,EMIN,S$\_$PR,ANG0,ANG1,SIPP)}

\noindent \underline{Input Arguments:}
\begin{description}
 \item[\tt INDF] is the fermion index (see \tbn{indf} and Figure 5 of
   \cite{Bardin:1999yd}). 
 \item[\tt ICUT] controls the kinds of cuts ({\em chain}) to be used.
   \begin{description}
     \item[{\tt ICUT} = --1:] (default) 
       allows for an $\smanp$ cut (a cut on $M^2_{\ff\fbf}$, the
         fermion and antifermion   
       invariant mass); the fastest branch based on \cite{Bardin:1991de}
     \item[{\tt ICUT} = 0:] {\bf not recommended}; branch is known to
       contain bugs. It
       allows for a cut on the acollinearity {\tt ACOL} of the \FF\ pair,
       on the minimum energy {\tt EMIN} of both fermion and antifermion,
       and for a geometrical acceptance cut
\cite{Bilenkii:1989zg}\footnote{As was shown recently 
       \cite{Christova:1999cc,Christova:1998tc,Christova:1999gh}, the old
       results of \cite{Bilenkii:1989zg} 
       contained bugs which occasionally didn't show up in comparisons
       as e.g. in \cite{Bardin:1995a2}. 
       The option is retained for
       back-compatibility with older versions only.}
     \item[{\tt ICUT} = 1:] $\smanp$ or $M^2_{\ff\fbf}$ cuts and geometrical 
       acceptance cut, based on \cite{Bardin:1991fu}
     \item[{\tt ICUT} = 2:]  new branch, replaces  {\tt ICUT} = 0 for
        realistic cuts {\tt ACOL} and {\tt EMIN}, based
       on~\cite{Christova:1999cc} 
     \item[{\tt ICUT} = 3:]   
       the same  branch, using {\tt ACOL} cut and  {\tt EMIN} cut
       but also with possibility to impose an additional acceptance
       cut \cite{Christova:1999gh}
   \end{description}
  \item[\tt ACOL] is the maximum acollinearity angle $ \xi^{\max} $ of the
    \FF\ pair in degrees ({\tt ICUT} = 0,2,3).
  \item[\tt EMIN] is the minimum energy $ E^{\min}_f $ of the fermion and
    antifermion in GeV ({\tt ICUT} = 0,2,3).
  \item[\tt S\_PR] is the minimum allowed invariant \FF\ mass $M^2_{\ff\fbf}$ 
    in GeV ({\tt ICUT} = --1,1) or, with some approximations, 
    the minimum allowed invariant mass of the 
    propagator after ISR\footnote{The invariant mass of the propagator
is not an observable quantity
unless specific assumptions on ISR and FSR are made.}
  \item[\tt ANG0] (default = $0^{\circ}$) is the minimum polar angle
    $\vartheta$ in degrees of the final-state antifermion.
  \item[\tt ANG1] (default = $180^{\circ}$) is the maximum polar angle
    $\vartheta$ in degrees of the final-state antifermion.
  \item[\tt SIPP] is a parameter  for cuts on the invariant mass of secondary
pairs for {\tt FSPP}=1,2 (see subsection 2.4). In older versions before v.6.30
the parameter {\tt SIPP} governed the calulation of corrections due to initial
state pairs and was recommended to be chosen equal to $\smanp$.
\end{description}

\subsection{Subroutine {\tt ZUINFO}\label{zuinfo}}
 
Subroutine {\tt ZUINFO} prints the values of  \zf\ flags and cuts.
 
\SUBR{CALL ZUINFO(MODE)}
 
\BS
\noindent \underline{Input Argument:}
\smallskip
 
\begin{description}
  \item[\tt MODE] controls the printing of \zf\ flag and cut values.
\begin{description}
  \item[{\tt MODE} = 0:] Prints all flag values.
  \item[{\tt MODE} = 1:] Prints all cut values.
\end{description}
\end{description}
 

\section{Interface routines of \zf\ \label{interfaces} }
\setcounter{equation}{0}
This Appendix describes technical details of the interface routines of
\zf.
Not all of them have been influenced by the updates over the years.
For the convenience of the user, we nevertheless decided to give a
complete overview and will repeat a substantial part of the material,
which was already presented in Section 4.3 of  \cite{Bardin:1999yd}.

Note that subroutine {\tt ZUWEAK} must be called prior to the
interfaces. 
As a consequence, the flags used in  {\tt ZUWEAK}
may influence the calculation of cross-sections and
asymmetries in the interfaces described now.
 
All subroutines need the following \underline{input arguments}:
\begin{description}
  \item[\tt INDF] is the fermion index (see \tbn{indf}).
  \item[\tt SQRS] is the centre-of-mass energy  \RS  ~in GeV.
  \item[\tt ZMASS] is the \Z\ mass  $\mzl$  in GeV.
\end{description}

We just mention that the interface using an S-matrix inspired language
is realized with the  {\tt SMATASY} package
\cite{Leike:1991pq,Riemann:1992gv,Kirsch:1995cf1,Kirsch:1995cf}. 
\subsection
{Subroutine {\tt ZUTHSM}
\label{zuthsm}}
 
Subroutine {\tt ZUTHSM} is used to calculate Standard Model
cross-sections and 
forward-back\-ward asymmetries. 
 
\SUBR{CALL ZUTHSM(INDF,SQRS,ZMASS,TMASS,HMASS,DAL5H,ALFAS,XS*,AFB*)}

 \BS
\noindent \underline{Input Arguments:}
\smallskip
\begin{description}
  \item[\tt TMASS] is the top quark mass  $\mtl$  in GeV, [10-400].
  \item[\tt HMASS] is the Higgs mass  $\mhl$  in GeV, [10-1000].
  \item[\tt DAL5H] is the value of $\dalhv$.
  \item[\tt ALFAS] is the value of the strong coupling constant
     $\alpha_s$  at $q^2 = \mzs$ (see also flag {\tt  QCDC} and
      factors {\tt QCDCOR}).
\end{description}
 
\nn \underline{Output Arguments}\footnote{An asterisk (*) following an
argument in a calling sequence is used to denote an output argument.}:
\smallskip
\begin{description}
  \item[\tt XS] is the total cross-section $\sigma_{\sss T}$  in nb.
  \item[\tt AFB] is the forward-backward asymmetry \afb .
\end{description}

\nn \underline{Output Internal Flag}:
\smallskip
\begin{description}
  \item[\tt INTRF=1]
\end{description}

\subsection
{Subroutine {\tt ZUATSM}
\label{zuatsm}}
 
Subroutine {\tt ZUATSM} is used to calculate 
differential cross-sections, $d \sigma/ d\cos \theta$,
in the Standard Model. 
 
\SUBR{CALL ZUATSM(INDF,SQRS,ZMASS,TMASS,HMASS,DAL5H,ALFAS,CSA*,DXS*)}

\BS
\noindent \underline{Input Arguments:}
\smallskip
\begin{description}
  \item[\tt TMASS] is the top quark mass  $\mtl$  in GeV, [10-400].
  \item[\tt HMASS] is the Higgs mass  $\mhl$  in GeV, [10-1000].
  \item[\tt ALQED] is the value of the running electromagnetic
                   coupling constant.
  \item[\tt ALFAS] is the value of the strong coupling constant
     $\alpha_s$  at $q^2 = \mzs $ (see factors {\tt QCDCOR}).
  \item[\tt CSA] is the cosine of the scattering angle.
\end{description}
 
\nn \underline{Output Arguments}:
\smallskip
\begin{description}
  \item[\tt DXS] is the theoretical differential cross-section.
\end{description}

\nn \underline{Output Internal Flag}:
\smallskip
\begin{description}
  \item[\tt INTRF=1]
\end{description} 
 
\subsection
{Subroutine {\tt ZUTPSM}
\label{zutpsm}}
 
Subroutine {\tt ZUTPSM} is used to calculate the 
tau
polarization and tau polarization asymmetry in the Standard Model. 
 
\SUBR{CALL ZUTPSM(SQRS,ZMASS,TMASS,HMASS,DAL5H,ALFAS,TAUPOL*,TAUAFB*)}

\BS
\noindent \underline{Input Arguments:}
\smallskip
\begin{description}
  \item[\tt HMASS] is the Higgs mass  $\mhl$  in GeV, [10-1000].
  \item[\tt DAL5H] is the value of $\dalhv$.
  \item[\tt ALFAS] is the value of the strong coupling constant
     $\alpha_s$  at $ q^2 = \mzs $ (see factors {\tt QCDCOR}).
\end{description}
 
\nn \underline{Output Arguments}:
\smallskip
\begin{description}
  \item[\tt TAUPOL] is the tau polarization  $A_{\mathrm{pol}}$ 
  of Equation 3.312 of \cite{Bardin:1999yd}.
  \item[\tt TAUAFB] is the tau polarization forward-backward
  asymmetry
  $A^{\mathrm{pol}}_{\sss FB}$  as defined in Equation 3.313 of
  \cite{Bardin:1999yd}. 
\end{description}
 
\nn \underline{Output Internal Flag}:
\smallskip
\begin{description}
  \item[\tt INTRF=1]
\end{description} 
 
\subsection
{Subroutine {\tt ZULRSM}
\label{zutlrm}}
 
Subroutine {\tt ZULRSM} is used to calculate the 
left-right asymmetry in the Standard Model. 
 
\SUBR{CALL ZULRSM(INDF,SQRS,ZMASS,TMASS,HMASS,DAL5H,ALFAS,POL,XSPL*,XSMI*)}
          
\BS
\noindent \underline{Input Arguments:}
\smallskip
\begin{description}
  \item[\tt TMASS] is the top quark mass  $\mtl$ in GeV, [40-300].
  \item[\tt HMASS] is the Higgs mass  $\mhl$ in GeV, [10-1000].
  \item[\tt DAL5H] is the value of $\dalhv$.
  \item[\tt ALFAS] is the value of the strong coupling constant
     $\alpha_s$ at $q^2 = \mzs $ (see also flag {\tt ALST}).
  \item[\tt POL] is the degree of longitudinal polarization of
                 electrons.
\end{description}
 
\nn \underline{Output Arguments}:
\smallskip
\begin{description}
  \item[\tt XSPL] is the cross-section for {\tt POL} $>$ 0
  \item[\tt XSMI] is the cross-section for {\tt POL} $<$ 0
\end{description}
 
\nn \underline{Output Internal Flag}:
\smallskip
\begin{description}
  \item[\tt INTRF=1]
\end{description} 

\subsection{Subroutine {\tt ZUXSA}\label{zuxsa}}
 
Subroutine {\tt ZUXSA} is used to calculate cross-section and
forward-backward asymmetry as described in Section 3.7.5 of
\cite{Bardin:1999yd}  as functions  of \RS, $\mzl$, $\gz$.
 
\SUBR{CALL ZUXSA(INDF,SQRS,ZMASS,GAMZ0,MODE,GVE,XE,GVF,XF,XS*,AFB*)}

\BS
\noindent \underline{Input Arguments:}
\begin{description}
  \item[\tt GAMZ0] is the total \Z\ width  $\gz$ in GeV.
  \item[\tt MODE] determines which weak couplings are used:
  \begin{description}
  \item[{\tt MODE} = 0:] {\tt XE} ({\tt XF}) are effective
    axial-vector couplings
     $\rab{\fe,\ff}$ for electrons (final-state fermions).
    \item[{\tt MODE} = 1:] {\tt XE} ({\tt XF}) are the effective weak
    neutral-current 
    amplitude normalizations  $\rhobi{\fe,\ff}$ for electrons (final-state
    fermions).
  \end{description}
  \item[\tt GVE] is the effective vector coupling for electrons
   $\rvab{\fe}$.
  \item[\tt XE] is  the effective axial-vector coupling  $\rab{\fe}$ or
    weak neutral-current amplitude normalization  $\rhobi{\fe}$
    for electrons (see {\tt MODE}).
  \item[\tt GVF] is the effective vector coupling for the final-state
    fermions  $\rvab{\ff}$.
  \item[\tt XF] is  the effective axial-vector coupling  $\rab{\ff}$ or
    the weak neutral-current amplitude normalization  $\rhobi{\ff}$
    for the final-state fermions (see {\tt MODE}).
\end{description}
 
\noindent \underline{Output Arguments:}
\smallskip
\begin{description}
  \item[\tt XS] is the cross-section $ \sigma_{\sss T}$ in nb.
  \item[\tt AFB] is the forward-backward asymmetry  \afb.
\end{description}
 
\nn \underline{Output Internal Flag}:
\smallskip
\begin{description}
  \item[\tt INTRF=3]
\end{description} 

\subsection{Subroutine {\tt ZUXSA2}\label{zuxsa2}}

Subroutine {\tt ZUXSA2} is used to calculate lepton cross-section and
forward-backward asymmetry as functions of \RS, $\mzl$, $\gz$, and of the weak
couplings {\em assuming lepton universality}. 
This routine is similar to {\tt ZUXSA} except that the couplings
are squared.
 
\SUBR{CALL ZUXSA2(INDF,SQRS,ZMASS,GAMZ0,MODE,GV2,X2,XS*,AFB*)}
 
\noindent \underline{Input Arguments:}
\begin{description}
  \item[\tt GAMZ0] is the total \Z\ width $\gz$ in GeV.
  \item[\tt MODE] determines which weak couplings are used:
  \begin{description}
    \item[{\tt MODE} = 0:] {\tt X2} is the square of the effective
    axial-vector coupling $\rab{\fl}$ for leptons.
    \item[{\tt MODE} = 1:] {\tt X2} is the square of the effective
neutral-current 
      amplitude normalization $\rhobi{\fl}$ for leptons.
  \end{description}
  \item[\tt GV2] is the square of the effective vector coupling
    $\rvab{\fl}$ for leptons.
  \item[\tt X2] is the square of the effective axial-vector coupling
    $\rab{\fl}$ or neutral-current amplitude normalization
    $\rhobi{\fl}$ for leptons (see {\tt MODE}).
\end{description}
 
\noindent \underline{Output Arguments:}
\begin{description}
  \item[\tt XS] is the cross-section $\sigma_{\sss T}$ in nb.
  \item[\tt AFB] is the forward-backward asymmetry \afb.
\end{description}
 
\nn \underline{Output Internal Flag}:
\begin{description}
  \item[\tt INTRF=4]
\end{description} 
 
\subsection{Subroutine {\tt ZUTAU}\label{zutau}}
 
Subroutine {\tt ZUTAU}  is used to calculate the $\tau^+$ polarization
as a function of \RS, $\mzl$, $\gz$, and the weak couplings.
 
\SUBR{CALL ZUTAU(SQRS,ZMASS,GAMZ0,MODE,GVE,XE,GVF,XF,TAUPOL*,TAUAFB*)}
 
\BS
\noindent \underline{Input Arguments:}
\smallskip
\begin{description}
  \item[\tt GAMZ0] is the total \Z\ width $\gz$ in GeV.
  \item[\tt MODE] determines which weak couplings are used:
  \begin{description}
    \item[{\tt MODE} = 0:] {\tt XE} ({\tt XF}) is the effective
    axial-vector coupling $\rab{\fe,\ff}$  for electrons (final-state
    fermions). 
    \item[{\tt MODE} = 1:] {\tt XE} ({\tt XF}) is the effective weak
neutral-current 
    amplitude normalization $\rhobi{\fe,\ff}$  for electrons (final-state
    fermions).
  \end{description}
  \item[\tt GVE] is the effective vector coupling for electrons
    $\rvab{\fe}$ .
  \item[\tt XE] is  the effective axial-vector coupling $\rab{\fe}$  or
    weak neutral-current amplitude normalization $\rhobi{\fe}$ 
    for electrons (see {\tt MODE}).
  \item[\tt GVF] is the effective vector coupling for the final-state
   fermions $\rvab{\ff}$ .
  \item[\tt XF] is  the effective axial-vector coupling $\rab{\ff}$  or
    weak neutral-current amplitude normalization $\rhobi{\ff}$ 
    for the final-state fermions (see {\tt MODE}).
\end{description}
 
\noindent \underline{Output Arguments:}
\begin{description}
  \item[\tt TAUPOL] is the $\tau$-lepton polarization $\lambda_{\tau}$
  \item[\tt TAUAFB] is the forward-backward asymmetry for polarized $\tau$-leptons $A^{\mathrm{pol}}_{\sss FB}$  
\end{description}

\nn \underline{Output Internal Flag}:
\begin{description}
  \item[\tt INTRF=3]
\end{description} 

\subsection{Subroutine {\tt ZUXSEC}\label{zuxsec}}
 
Subroutine {\tt ZUXSEC} is  used to calculate the cross
section as a function of \RS, $\mzl$, $\gz$, $\Gamma_e$ and $\Gamma_f$.
 
\SUBR{CALL ZUXSEC(INDF,SQRS,ZMASS,GAMZ0,GAMEE,GAMFF,XS*)}
 
\noindent \underline{Input Arguments:}
\begin{description}
  \item[\tt GAMZ0] is the total \Z\ width $\gz$  in GeV.
  \item[\tt GAMEE] is the partial \Z\ decay width $\Gamma_e$  in GeV.
  \item[\tt GAMFF] is the partial \Z\ decay width $\Gamma_f$  in GeV;
  if {\tt INDF}=10, {\tt GAMFF}=$\Gamma_{h}$.
\end{description}
 
\nn \underline{Output Internal Flag}:
\smallskip
\begin{description}
  \item[\tt INTRF=2]
\end{description} 

\subsection{Subroutine {\tt ZUXAFB}\label{zulr}}

Subroutine {\tt ZUXAFB} is  used to calculate the cross
section as a function of \RS, $\mzl$, $\gz$, $\Gamma_e$ and $\Gamma_f$.
 
\SUBR{CALL ZUXAFB(INDF,SQRS,ZMASS,GAMZ0,PFOUR,PVAE2,PVAF2,XS*,AFB*)}
 
\noindent \underline{Input Arguments:}
\begin{description}
  \item[\tt GAMZ0] is the total \Z\ width $\gz$  in GeV.
  \item[\tt PFOUR] is the product of vector and axial-vector couplings
                   $\four$.
  \item[\tt PVAE2] is $\vaeII$.
  \item[\tt PVAF2] is $\vafII$.
\end{description}
 
\noindent \underline{Output Argument:}
\begin{description}
  \item[\tt XS] is the cross-section $\sigma_{\sss T}$ in nb.
  \item[\tt AFB] is the forward-backward asymmetry \afb.
\end{description}
 
\nn \underline{Output Internal Flag}:
\smallskip
\begin{description}
  \item[\tt INTRF=5]
\end{description} 

\subsection{Subroutine {\tt ZUALR}
\label{zualr}}

Subroutine {\tt ZUALR} is reserved for the fit of $A_{\sss LR}$.
     
\SUBR{CALL ZUALR(SQRS,ZMASS,GAMZ0,MODE,GVE,XE,GVF,XF,TAUPOL*,TAUAFB*)}

\nn \underline{Output Internal Flag}:
\smallskip
\begin{description}
  \item[\tt INTRF=6]
\end{description} 
\subsection{Subroutine {\tt ZVWEAK}}\label{zvweak}
The strength of  the Wtb vertex, $|V_{tb}|$, can be changed by using
the subroutine {\tt ZVWEAK} instead of {\tt ZUWEAK}; see \sect{intro}.

\SUBR{CALL ZVWEAK(ZMASS,TMASS,HMASS,DAL5H,V$\_$TB,ALFAS)}

\noindent \underline{Input Arguments:}
\smallskip
\begin{description}
  \item[\tt V\_TB] is the value of $|V_{tb}|$; the Standard Model calculation of {\tt ZUWEAK} corresponds to
$|V_{tb}|=1$.
\end{description}
{\tt ZVWEAK} has to be complemented with new interface routines, all now with the the additional input argument {\tt V\_TB}:
\\
{\tt ZVTHSM(INDF,SQRS,ZMASS,TMASS,HMASS,DAL5H,V\_TB,ALFAS,XS*,AFB*)} \\
{\tt ZVTPSM(SQRS,ZMASS,TMASS,HMASS,DAL5H,V\_TB,ALFAS,TAUPOL*,TAUAFB*)} \\
{\tt ZVLRSM(INDF,SQRS,ZMASS,TMASS,HMASS,DAL5H,V\_TB,ALFAS,POL,XSPL*,XSMI*)} \\
{\tt ZVATSM(INDF,SQRS,ZMASS,TMASS,HMASS,DAL5H,V\_TB,ALFAS,CSA,DXS*)} \\

These subroutines replace {\tt ZUTHSM}, {\tt ZUTPSM}, {\tt ZULRSM} and {\tt ZUATSM}.

\subsection{Subroutine {\tt ZU\_APV}
\label{zuapv}}

With subroutine {\tt ZU\_APV} 
measurements of the weak charge, $Q_W$, (see \sect{sect:apv})
can be included into the global tests of the Standard Model.

\SUBR{CALL ZU\_APV(ZMASS,TMASS,HMASS,SIN2TW,UMASS,DMASS,C1U*,C1D*,C2U*,C2D*)}
 
\noindent \underline{Input Arguments:}
\begin{description}
  \item[\tt SIN2TW] is the $\sin$ of the weak mixing angle.
  \item[\tt UMASS] is the u-quark mass (constituent).
  \item[\tt DMASS] is the d-quark mass (constituent).
\end{description}
 
\noindent \underline{Output Argument:}
\begin{description}
  \item[\tt C1U] -- coupling parameter, $C_{1u}=2a_e v_u$ 
  \item[\tt C1D] -- coupling parameter, $C_{1d}=2a_e v_d$ 
  \item[\tt C2U] -- coupling parameter, $C_{2u}=2v_e a_u$  
  \item[\tt C2D] -- coupling parameter, $C_{2d}=2v_e a_d$ 
\end{description}


\eqnzero
\newpage


\begin{table}[bthp]
{\fontsize{9pt}{10pt}\selectfont
\begin{tabular}{|l|cccccccccccc|}
\hline
Final- & & & & & & & & & & & &  \\
state  & $\nu{\bar{\nu}}$&$e^+e^-$&$\mu^+\mu^-$&
          $\tau^+\tau^-$&$u\bar{u}$&$d\bar{d}$& 
          $c\bar{c}$&$s\bar{s}$&$t\bar{t}$&
          $b\bar{b}$&\small{hadrons}&\small{Bhabha}\\
fermions& & & & & & & & & & & &  \\ \hline  \hline
\hline{\tt INDF}  &0&1&2&3&4&5&6&7&8&
         9&10&11\\ \hline
\end{tabular}
}
\vspace*{0.3cm}

\caption[Indices for the selection of final states]
{\it
Indices for the selection of final states.
Note that \mbox{{INDF=0}} returns values for one neutrino species,
while \mbox{{INDF=10}} returns values for the five-flavour inclusive (udscb)
hadronic channel. Also note that \mbox{ {\tt INDF} = 0,1} includes
only s-channel calculations while \mbox{{\tt INDF} = 8} always returns
zero.
\label{indf}
}
\end{table}

\bigskip

\clearpage

\begin{table}[bthp]\centering
\renewcommand{\arraystretch}{1}
{\fontsize{9pt}{10pt}\selectfont
\begin{tabular}{|c|c|c|c|c|c|}  
\hline
I  &$~~~${\tt 'FLAG'}$~~~$& $~~$ name in $~~$ & Position in {\tt DIZET}& Position in {\tt ZFITTER}& default \\
     &          & programs  &{\tt NPAR(1:25)}&{\tt NPAR(1:30)} & value \\
\hline
\hline
 1 &{\tt AFBC} &{\tt IAFB}  &    & 13 & 1 \\
 2 &{\tt SCAL} &{\tt ISCAL} &  9 & 15 & 0 \\
 3 &{\tt SCRE} &{\tt ISCRE} &  6 &    & 0 \\
 4 &{\tt AMT4} &{\tt IAMT4} &  2 & 16 & 4 \\
 5 &{\tt BORN} &{\tt IBORN} &    & 14 & 0 \\
 6 &{\tt BOXD} &{\tt IBOX}  &    &  4 & 1 \\
 7 &{\tt CONV} &            &    &    & 1 \\
 8 &{\tt FINR} &{\tt IFINAL}&    &  9 & 1 \\
 9 &{\tt FOT2} &{\tt IPHOT2}&    & 10 & 3 \\
10 &{\tt GAMS} &            &    &  5 & 1 \\
11 &{\tt DIAG} &            &    &  7 & 1 \\
12 &{\tt INTF} &{\tt INTERF}&    &  8 & 1 \\
13 &{\tt BARB} &{\tt IBARB} & 10 &    & 2 \\
14 &{\tt PART} &{\tt IPART} &    & 17 & 0 \\
15 &{\tt POWR} &            &    &    & 1 \\
16 &{\tt PRNT} &            &    &    & 0 \\
17 &{\tt ALEM} &{\tt IALEM} &  7 & 20 & 3 \\
18 &{\tt QCDC} &{\tt IQCD}  &  3 &  3 & 3 \\
19 &{\tt VPOL} &{\tt IHVP}  &  1 &  2 & 1 \\
20 &{\tt WEAK} &{\tt IWEAK} &    &  1 & 1 \\
21 &{\tt FTJR} &{\tt IFTJR} & 11 &    & 1 \\
22 &{\tt EXPR} &{\tt IFACR} & 12 &    & 0 \\
23 &{\tt EXPF} &{\tt IFACT} & 13 & 19 & 0 \\
24 &{\tt HIGS} &{\tt IHIGS} & 14 &    & 0 \\
25 &{\tt AFMT} &{\tt IAFMT} & 15 &    & 3 \\
26 &{\tt CZAK} &{\tt ICZAK} & 17 &    & 1 \\
27 &{\tt PREC} &{\tt NPREC} &    &    &10 \\
28 &{\tt HIG2} &{\tt IHIG2} & 18 &    & 0 \\
29 &{\tt ALE2} &{\tt IALE2} & 19 & 21 & 3 \\
30 &{\tt GFER} &{\tt IGFER} & 20 &    & 2 \\
31 &{\tt ISPP} &{\tt ISRPPR}&    &    & 2 \\
32 &{\tt FSRS} &            &    &    & 1 \\
33 &{\tt MISC} &{\tt IMISC} &    &    & 0 \\
34 &{\tt MISD} &{\tt IMISD} &    &    & 1 \\
35 &{\tt IPFC} &            &    &    & 5 \\
36 &{\tt IPSC} &            &    &    & 0 \\
37 &{\tt IPTO} &            &    &    & 3 \\
38 &{\tt FBHO} &            &    &    & 0 \\
39 &{\tt FSPP} &{\tt IFSPPR}&    &    & 0 \\
40 &{\tt FUNA} &{\tt IFUNAN}&    &    & 0 \\
41 &{\tt ASCR} &            &    &    & 1 \\
42 &{\tt SFSR} &{\tt ISFSR} & 23 &    & 1 \\
43 &{\tt ENUE} &            &    &    & 1 \\
44 &{\tt TUPV} &{\tt ITUPV} &    &    & 1 \\
45 &{\tt DMWW} &{\tt IDMWW} & 24 &    & 0 \\
46 &{\tt DSWW} &{\tt IDSWW} & 25 &    & 0 \\
   &           &{\tt IMOMS} &  4 &    & 1 \\
   &           &{\tt IMASS} &  5 &    & 0 \\
   &           &{\tt IMASK} &  8 &    & 0 \\
   &           &{\tt IEWLC} & 16 &    & 1 \\
   &           &{\tt IDDZZ} & 21 &    & 1 \\
   &           &{\tt IAMW2} & 22 &    & 0 \\
\hline
\end{tabular}
}
\vspace*{0.2cm}

\caption[Flags used in {\tt DIZET} and {\tt ZFITTER}] {\it Flag
settings for \zf\ and {\tt DIZET}; the flags are listed in the order
of vector {\tt IFLAGS}.  The corresponding names used internally in
the programs, the position of the flags in vector {\tt NPAR(1:25)} of
{\tt DIZET} (called {\tt NPARD} in \zf) and {\tt NPAR(1:30)} of {\tt
ZFITTER} and the default values are given }
\label{tab:xxxx}
\end{table}

\clearpage
\providecommand{\href}[2]{#2}
\end{document}